\PassOptionsToPackage{unicode}{hyperref}
\PassOptionsToPackage{hyphens}{url}
\PassOptionsToPackage{dvipsnames,svgnames,x11names}{xcolor}
\documentclass[
  12pt]{article}
\usepackage{multirow}
\usepackage{amsmath,amssymb}
\usepackage{iftex}
\ifPDFTeX
  \usepackage[T1]{fontenc}
  \usepackage[utf8]{inputenc}
  \usepackage{textcomp} 
\else 
  \usepackage{unicode-math}
  \defaultfontfeatures{Scale=MatchLowercase}
  \defaultfontfeatures[\rmfamily]{Ligatures=TeX,Scale=1}
\fi
\usepackage{lmodern}
\ifPDFTeX\else  
\fi
\IfFileExists{upquote.sty}{\usepackage{upquote}}{}
\IfFileExists{microtype.sty}{
  \usepackage[]{microtype}
  \UseMicrotypeSet[protrusion]{basicmath} 
}{}
\makeatletter
\@ifundefined{KOMAClassName}{
  \IfFileExists{parskip.sty}{%
    \usepackage{parskip}
  }{
    \setlength{\parindent}{0pt}
    \setlength{\parskip}{6pt plus 2pt minus 1pt}}
}{
  \KOMAoptions{parskip=half}}
\makeatother
\usepackage{xcolor}
\makeatletter
\ifx\paragraph\undefined\else
  \let\oldparagraph\paragraph
  \renewcommand{\paragraph}{
    \@ifstar
      \xxxParagraphStar
      \xxxParagraphNoStar
  }
  \newcommand{\xxxParagraphStar}[1]{\oldparagraph*{#1}\mbox{}}
  \newcommand{\xxxParagraphNoStar}[1]{\oldparagraph{#1}\mbox{}}
\fi
\ifx\subparagraph\undefined\else
  \let\oldsubparagraph\subparagraph
  \renewcommand{\subparagraph}{
    \@ifstar
      \xxxSubParagraphStar
      \xxxSubParagraphNoStar
  }
  \newcommand{\xxxSubParagraphStar}[1]{\oldsubparagraph*{#1}\mbox{}}
  \newcommand{\xxxSubParagraphNoStar}[1]{\oldsubparagraph{#1}\mbox{}}
\fi
\makeatother

\usepackage{longtable,booktabs,array}
\usepackage{calc} 
\usepackage{etoolbox}
\makeatletter
\patchcmd\longtable{\par}{\if@noskipsec\mbox{}\fi\par}{}{}
\makeatother
\IfFileExists{footnotehyper.sty}{\usepackage{footnotehyper}}{\usepackage{footnote}}
\makesavenoteenv{longtable}
\usepackage{graphicx}
\makeatletter
\def\maxwidth{\ifdim\Gin@nat@width>\linewidth\linewidth\else\Gin@nat@width\fi}
\def\maxheight{\ifdim\Gin@nat@height>\textheight\textheight\else\Gin@nat@height\fi}
\makeatother
\setkeys{Gin}{width=\maxwidth,height=\maxheight,keepaspectratio}
\makeatletter
\def\fps@figure{htbp}
\makeatother

\makeatletter
\@ifpackageloaded{caption}{}{\usepackage{caption}}
\AtBeginDocument{%
\ifdefined\contentsname
  \renewcommand*\contentsname{Table of contents}
\else
  \newcommand\contentsname{Table of contents}
\fi
\ifdefined\listfigurename
  \renewcommand*\listfigurename{List of Figures}
\else
  \newcommand\listfigurename{List of Figures}
\fi
\ifdefined\listtablename
  \renewcommand*\listtablename{List of Tables}
\else
  \newcommand\listtablename{List of Tables}
\fi
\ifdefined\figurename
  \renewcommand*\figurename{Figure}
\else
  \newcommand\figurename{Figure}
\fi
\ifdefined\tablename
  \renewcommand*\tablename{Table}
\else
  \newcommand\tablename{Table}
\fi
}
\@ifpackageloaded{float}{}{\usepackage{float}}
\floatstyle{ruled}
\@ifundefined{c@chapter}{\newfloat{codelisting}{h}{lop}}{\newfloat{codelisting}{h}{lop}[chapter]}
\floatname{codelisting}{Listing}

\makeatother
\makeatletter
\@ifpackageloaded{caption}{}{\usepackage{caption}}
\@ifpackageloaded{subcaption}{}{\usepackage{subcaption}}
\makeatother

\ifLuaTeX
  \usepackage{selnolig}  
\fi
\usepackage[]{natbib}
\usepackage{bookmark}

\IfFileExists{xurl.sty}{\usepackage{xurl}}{} 
\hypersetup{
  pdftitle={Title},
  pdfauthor={Author 1; Author 2},
  pdfkeywords={3 to 6 keywords, that do not appear in the title},
  colorlinks=true,
  linkcolor={blue},
  filecolor={Maroon},
  citecolor={Blue},
  urlcolor={Blue},
  pdfcreator={LaTeX via pandoc}}

\newcommand{\anon}{1}

\usepackage{authblk}
\usepackage{threeparttable}

\date{}

\begin{document}

\def\spacingset#1{\renewcommand{\baselinestretch}%
{#1}\small\normalsize} \spacingset{1}


\renewcommand\Authfont{\normalsize}
\renewcommand\Affilfont{\small}
\setlength{\affilsep}{6pt}

\if1\anon
{
  \title{\bf A Workflow for Evaluating Regional Treatment Effect
  Heterogeneity in Multi-Regional Clinical Trials}

  \author[1]{Cong Zhang}
  \author[2]{Meihua Long}
  \author[3]{Tianyu Zheng}
  \author[4]{Konstantinos Sechidis}
  \author[1]{Xiaoni Liu}
  \author[5]{Sophie Sun}
  \author[5]{Yao Chen}
  \author[6]{Xinyi Zhang}
  \author[7]{Shuhei Kaneko}
  \author[4,$\ast$]{Björn Bornkamp}
  \author[2,8$\ast$]{Yan Hou}

  \affil[1]{\parbox[t]{0.85\textwidth}{\small China Novartis
    Institutes for BioMedical Research Co., Shanghai, China}}
  \affil[2]{\parbox[t]{0.85\textwidth}{\small Department of Biostatistics, School of Public Health, Peking University, Beijing, China}}
  \affil[3]{\small Peking University Cancer Hospital, Beijing, China}
  \affil[4]{\parbox[t]{0.85\textwidth}{\small Advanced Methodology
    and Data Science, Novartis Pharma AG, Basel, Switzerland}}
  \affil[5]{\parbox[t]{0.85\textwidth}{\small Advanced Methodology
    and Data Science, Novartis Pharmaceuticals Corporation,
    East Hanover, NJ, USA}}
  \affil[6]{\small Department of Public Health Sciences, University of Chicago, IL, USA}
  \affil[7]{\small Biostatistics CRM/NS/IMM, Advanced Quantitative Sciences, Global Drug Development Division, Novartis Pharma K.K., Tokyo, Japan.}
  \affil[8]{\small Key Laboratory of Carcinogenesis and Translational Research (Ministry of Education), Peking University Cancer Hospital \& Institute, Beijing, China}
  \affil[$\ast$]{\small Co-corresponding authors.
    Email: bjoern.bornkamp@novartis.com; houyan@bjmu.edu.cn}

  \maketitle
} \fi

\if0\anon
{
  \bigskip
  \bigskip
  \bigskip
  \begin{center}
    {\LARGE\bf A Workflow for Evaluating Regional Treatment Effect Heterogeneity in Multi-Regional Clinical Trials}
\end{center}
  \medskip
} \fi

\bigskip
\begin{abstract}
Multi-regional clinical trials (MRCTs) enable efficient global drug development by assessing treatment effects across regions within a single protocol. While powered for overall efficacy, MRCTs are typically not designed to provide confirmatory evidence
on regional differences, making an assessment of observed regional heterogeneity largely exploratory and susceptible to sampling variability. Despite this challenge, understanding regional heterogeneity remains important for interpretation and regulatory decision-making.
This paper proposes a structured, question-driven framework to guide exploratory assessments of regional heterogeneity in MRCTs. We formulate four key questions to clarify the objectives of such analyses and propose a set of statistical methods to address them. Simulation studies evaluate performance under scenarios with no heterogeneity and heterogeneity driven by observed or unobserved treatment effect modifiers, illustrating how a structured approach can support transparent and cautious interpretation.
\end{abstract}

\noindent%
\vfill

\newpage
\spacingset{1.8} 

\section{Introduction}\label{sec1}
Multi-regional clinical trials (MRCTs) generate evidence on treatment efficacy and safety across multiple regions under a single global protocol. As outlined in the ICH E17 guideline \citep{ICH2017}, MRCTs improve the efficiency of drug development by enabling simultaneous regulatory submissions, reducing duplication of regional studies, and allowing earlier worldwide access to effective therapies. 

Although MRCTs follow a unified protocol, it is often of interest to examine whether treatment effects vary across regions and to understand whether systematic reasons underlie any observed heterogeneity. Such heterogeneity may arise from intrinsic factors, including genetics, age, sex, body size, and organ function, or from extrinsic factors related to environmental, cultural, or healthcare-system differences~\citep{yusuf2016interpreting}.

The sample size for an MRCT is typically determined to ensure sufficient power for evaluating the overall treatment effect, under the assumption that this effect is uniform across the entire target population and thus all participating regions~\citep{ICH2017}. Consequently, regional treatment effect estimates in MRCTs typically lack the precision needed for confirmatory conclusions: observed differences in treatment effects across regions may be driven by sampling variability alone and may not reflect true underlying differences. 

Assessments and data-driven explanations of regional heterogeneity are therefore inherently exploratory, requiring cautious interpretation supported by external evidence~\citep{Yusuf1991}. Incorporating such information, for example, on factors known to differ across regions or expected to be prognostic or predictive of the outcome, is essential for meaningful interpretation. Regional regulatory authorities have
further emphasized the importance of structured investigation when regional inconsistencies are observed.

A range of statistical methods have been proposed for assessing regional 
consistency, including visual tools such as forest plots, formal tests 
such as treatment-by-region interaction tests 
\citep{li2021simultaneous}, the Japanese Ministry of Health, Labour and Welfare (MHLW) consistency criteria
\citep{ministry2007basic}, and outlier detection techniques 
\citep{viechtbauer2010outlier}. These methods typically focus on detecting regional differences rather than explaining them. Yet region is not itself a direct effect modifier but a composite proxy for observed and unobserved intrinsic and extrinsic factors, so understanding regional heterogeneity requires looking beyond region-level summaries to the underlying patient characteristics.Moreover, analyses of regional heterogeneity are often conducted post hoc. This introduces analytic flexibility and researcher degrees of freedom, which may be exercised based on personal experience and preferences, potentially increasing variability of results and reducing their quality \citep{Silberzahn2018}.

To address this concern, we propose a structured approach for assessing and potentially explaining observed regional heterogeneity in MRCTs, organized around four guiding questions. It progresses from detecting heterogeneity, through identifying potential region-associated effect modifiers, to quantifying the extent to which observed covariates explain the regional differences. 

The remainder of this article is organized as follows. Section 2 presents the four key questions for structuring the assessment of regional heterogeneity. Section 3 describes one approach for answering these questions using a specific set of statistical methods. Section 4 evaluates these methods in a comprehensive simulation study spanning three scenarios: no regional heterogeneity, and regional heterogeneity driven by observed and unobserved effect modifiers, respectively. Section 5 discusses the implications of our findings and directions for future research.

\section{Framework for Structured Regional Heterogeneity Analysis}\label{sec:framework}

ICH E17 recommends that the statistical analysis plan for an MRCT include an assessment of the consistency of treatment effects across regions and key subpopulations, where consistency is understood as the absence of clinically relevant differences. Typical components of such an assessment include descriptive summaries, graphical displays (e.g., forest plots), covariate-adjusted model-based estimates, and exploratory evaluations of treatment-by-region interactions.

When clinically relevant regional differences are observed, ICH E17 further advises structured post-hoc analysis to investigate whether these discrepancies can be plausibly explained by imbalances in intrinsic and extrinsic factors (including prognostic and potentially predictive variables), and to incorporate additional evidence when discrepancies remain unexplained. Prior literature and case studies similarly advocate for an exploratory, structured approach that first examines known sources of heterogeneity and regional imbalances and then probes unexpected differences using supportive data and external evidence when needed.

Motivated by these principles, we propose a set of guiding questions for the post-hoc assessment of region-associated treatment effect heterogeneity (TEH) in an MRCT setting.

\subsection{Notation and conceptual framework for regional heterogeneity}
\label{sec:conceptual}

When treatment effects appear to differ across regions, it is of interest to assess whether the observed heterogeneity can be explained by differences in intrinsic and extrinsic factors. In this section, we present a conceptual framework for characterizing regional heterogeneity.

Let $Region$ denote a regional grouping in the MRCT, and let $\mathbf{X}=(X_1,\ldots,X_p)$ denote a vector of observed baseline covariates capturing intrinsic and extrinsic factors. We further define an unobserved component $U$ representing factors not captured in $\mathbf{X}$ but potentially related to both region and treatment response. Throughout, the term ``treatment effect'' denotes a pre-specified contrast between outcomes under treatment assignment $Z=1$ and $Z=0$ on an endpoint-appropriate scale, and our interest lies in how the distribution of this treatment effect may vary across regions.

\begin{figure*}
    \centerline{\includegraphics[width=0.4\linewidth]{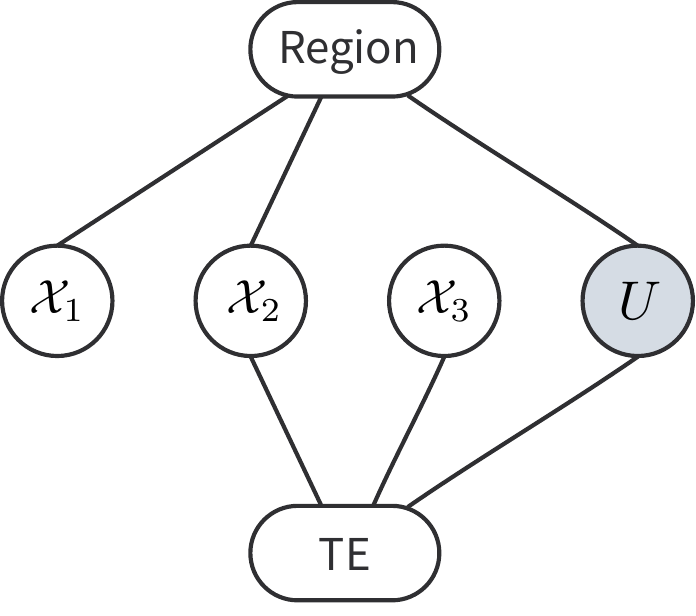}}
    \caption{\textbf{Conceptual association graph for structured regional heterogeneity analysis.}
    Nodes represent the pre-defined regional grouping ($Region$), observed baseline covariates partitioned into three subsets $\mathcal{X}_1, \mathcal{X}_2, \mathcal{X}_3$, unobserved factors $U$ (shaded), and the treatment effect $TE$. Edges represent associations; no causal direction is asserted.
    The graph encodes the roles of different covariate types.
    $\mathcal{X}_1$: covariates that differ across regions but do not modify the treatment effect (connected to $Region$ only).
    $\mathcal{X}_2$: \emph{region-associated effect modifiers}, covariates that both modify the treatment effect and exhibit meaningful imbalance across regions.
    $\mathcal{X}_3$: covariates that modify the treatment effect but are balanced across regions (connected to $TE$ only).
    $U$ represents unobserved factors associated with both $Region$ and $TE$;
    the pathway $Region$--$U$--$TE$ gives rise to residual region-associated heterogeneity, with $Region$ potentially acting as a surrogate for these unmeasured factors. \label{fig1} }
\end{figure*}

To elucidate the mechanisms behind regional heterogeneity, Figure~\ref{fig1} summarizes a conceptual framework linking region, observed covariates, unobserved factors, and treatment effect heterogeneity. Within this framework, baseline factors are classified into four groups according to their relationship with region and the treatment effect. We use calligraphic symbols $\mathcal{X}_i$ to denote these conceptual classes of covariates (subsets of $\mathbf{X}$), whereas $X_j$ denotes an individual observed covariate:

\begin{itemize}
    \item \(\mathcal{X}_1\): covariates that are unevenly distributed across regions but do not modify the treatment effect.
    \item \(\mathcal{X}_2\): covariates termed region-associated effect modifiers, i.e., covariates that both modify the treatment effect and exhibit meaningful imbalance across regions.
    \item \(\mathcal{X}_3\): covariates that modify the treatment effect but are not associated with region.
    \item \(U\): unobserved factors that may drive residual regional heterogeneity; in particular, they may be region-associated and influence the treatment effect.
\end{itemize}

The key to explaining regional heterogeneity in treatment effects are covariates in $\mathcal{X}_2$, the region-associated effect modifiers, which satisfy two conditions simultaneously: (i) they exhibit regional imbalance and (ii) they modify the treatment effect. When such covariates are absent from the measured set (or insufficient to account for observed patterns), residual region-associated heterogeneity may indicate the influence of unobserved factors $U$, with region acting as a proxy.

\subsection{Guiding questions for regional heterogeneity exploration}

Based on this conceptual framework, we propose that a regional heterogeneity assessment should address the following four analytical questions:
\begin{itemize}
    \item[(Q1)] Does regional heterogeneity in treatment effects exist?
    \item[(Q2)] Which baseline covariates are unevenly distributed across regions?
    \item[(Q3)] Which baseline covariates modify the treatment effect?
    \item[(Q4)] How do treatment effects vary along key effect modifiers across regions?
\end{itemize}

Q1 addresses the overarching question of whether regional heterogeneity exists, given the chosen categorization of the region covariate. Interaction tests are commonly used to answer this question. Importantly, the resulting p-values should not be interpreted as binary decision rules but rather on a continuous scale \cite{cole2020}: the trial was not designed for this comparison, and exploratory investigations do not lend themselves to simple binary decision-making.

Q2 concerns which covariates are imbalanced across regions. ICH E17 places considerable emphasis on characterizing such imbalances in terms of baseline covariates. One approach is to fit a predictive model with region as the outcome variable, then report the evidence against regional homogeneity across all baseline covariates. When there is some evidence against homogeneity, important predictors of region can be identified (for example, by assessing variable importances from the model). In practice, differences across regions in terms of baseline covariates can often be characterized with reasonable reliability. The covariates identified in this way may belong to $\mathcal{X}_1$ or $\mathcal{X}_2$.

Q3 addresses whether treatment effects vary as a function of baseline covariates. This question is well known to be very challenging to answer reliably given the sample sizes available in a single trial or even a small number of trials \cite{Sechidis2025WATCH}. As with Q2, a model-based approach can be employed: one first assesses the evidence against treatment effect homogeneity across all baseline covariates (for example, using a global interaction test), and when there is some evidence against homogeneity, suggest important predictors of the treatment effect (for example, via variable importances). The covariates identified in this way may belong to $\mathcal{X}_2$ or $\mathcal{X}_3$. 

Q4 asks how treatment effects vary along key effect modifiers across regions, and specifically whether the intersection of region-associated covariates (from Q2) and treatment effect modifiers (from Q3) can plausibly account for the observed regional heterogeneity. When covariates appear in both ranked lists, they are candidates for $\mathcal{X}_2$. Descriptive displays can then be used to visualize. For example, one might plot the treatment effect modifier on the x-axis and the treatment effect on the y-axis, with histograms highlighting the distribution of the covariate across regions. These displays complement the results from Q1–Q3 and facilitate discussion with multidisciplinary experts.

\subsection{Decision roadmap for workflow interpretation and recommended actions}
\label{sec:roadmap}

While the proposed workflow is inherently exploratory and findings must be interpreted in conjunction with clinical judgment and external evidence, it is useful to outline a structured decision roadmap linking the outcome at each analytical step to recommended next actions. This roadmap is not intended as a set of rigid decision rules, but rather as a navigational guide that supports multidisciplinary teams in determining the depth of investigation warranted and in framing conclusions appropriate to the level of evidence. Table~\ref{tab:roadmap} provides a structured summary of this decision logic, linking each assessment node to its possible outcomes and recommended actions.

\begin{table}[!htbp]
\footnotesize
\renewcommand{\arraystretch}{0.9} 
\setlength{\tabcolsep}{3pt}       
\centering
\caption{Decision roadmap: decision nodes, outcomes, and recommended actions.}
\label{tab:roadmap}
\begin{tabular}{p{2.8cm} p{3.2cm} p{6.0cm} p{3.0cm}}
\toprule
\textbf{Decision Node} &
\textbf{Possible Outcome} &
\textbf{Recommended Action} &
\textbf{Terminal Conclusion} \\
\midrule
\multirow{3}{2.8cm}{Node 1, answer Q1: Regional TEH?}
  & No evidence
  & Document regional consistency; no further decomposition needed.
  & \textbf{T1:} Regional consistency \\
\cmidrule(l){2-4}
  & Low evidence
  & Proceed cautiously to Questions~2--3; document ambiguity; interpret subsequent findings conservatively.
  & --- \\
\cmidrule(l){2-4}
  & Strong evidence
  & Proceed to Questions~2--3 to investigate potential sources.
  & --- \\
\midrule
\multirow{2}{2.8cm}{Node 2, answer Q2: Covariates imbalanced across regions?}
  & Imbalanced covariates identified
  & Record ranked list of region-associated covariates; carry forward to overlap assessment.
  & --- \\
\cmidrule(l){2-4}
  & No meaningful imbalances
  & Regional populations are similar on measured covariates; heterogeneity unlikely driven by observed population differences.
  & --- \\
\midrule
\multirow{2}{2.8cm}{Node 3, answer Q3: Effect modifiers identified?}
  & Clear effect modifiers emerge
  & Record ranked list of effect-modifying covariates; carry forward to overlap assessment.
  & --- \\
\cmidrule(l){2-4}
  & No clear effect modifiers
  & Treatment effect heterogeneity not attributable to measured covariates; consider optional extensions (Sec.~3.6).
  & \textbf{T4:} Unexplained heterogeneity; \textbf{T5:} Sampling variability \\
\midrule
\multirow{2}{2.8cm}{Node 4, overlap of Q2 \& Q3: Regional covariates $\cap$ effect modifiers?}
  & Non-empty overlap
  & Measured covariates plausibly explain regional pattern; proceed to Q4 for visualization and synthesis.
  & \textbf{T2:} Identified candidate factor; \textbf{T3:} Partial explanation \\
\cmidrule(l){2-4}
  & Empty overlap
  & Region may act as surrogate for unmeasured factors; document unexplained component.
  & \textbf{T4:} Unexplained heterogeneity \\
\bottomrule
\end{tabular}
\end{table}

\subsubsection*{Summary of terminal conclusions}

Across the nodes above, the workflow converges on a limited set of terminal conclusions for the multidisciplinary team:
\begin{enumerate}
    \item[\textbf{T1.}] \textbf{Report regional consistency:} No extensive further exploration required.

    \item[\textbf{T2.}] \textbf{Report heterogeneity with an identified candidate factor:} supported by a coherent chain of evidence from Nodes~1--4, with explicit acknowledgment of the exploratory nature and absence of formal causal claims.

    \item[\textbf{T3.}] \textbf{Report heterogeneity with partial explanation:} some factors identified but residual variability remains; implications for generalizability (and labeling) should be discussed.

    \item[\textbf{T4.}] \textbf{Report unexplained heterogeneity:} no measured covariate adequately accounts for the observed differences; possible attribution to unmeasured factors should be discussed, and planning implications for future trials noted.

    \item[\textbf{T5.}] \textbf{Attribute observed variation primarily to sampling variability:} supported by small sample sizes, absence of any systematic pattern, and robustness to shrinkage or sensitivity analyses.
\end{enumerate}

In practice, which branch to take at each node of the roadmap, and thus which terminal conclusion to arrive at, may not be entirely clear and will depend on the context. The decision should be guided by the outputs resulting from the analyses described in more detail below, as well as by clinical plausibility and external
evidence. The roadmap is intended to structure this exploratory investigation and ensure that the depth of inquiry is proportionate to the strength and consistency of the signals observed, consistent with the principles articulated in ICH~E17~\citep{ICH2017} and the general philosophy of structured exploratory analysis~\citep{bretz2023role, baillie2023good}.

\section{Answering the Four Questions --- Methods and Illustration}\label{methods-example}

In this section, we describe one implementation of the question-driven framework
introduced in Section~\ref{sec:framework} and illustrate it on two simulated
datasets. The framework itself is agnostic to the choice of statistical methods:
any approach that provides principled answers to Q1--Q4 can be used. For example,
the PLATO trial analysis \citep{CarrollFleming2013PLATO} implicitly followed a similar logic,
using univariate Cox regression to identify covariates that were simultaneously
imbalanced across regions and interacted with treatment. Here, we employ an
implementation based on doubly robust (DR) pseudo-outcomes combined with
conditional random forests (CRFs) for variable ranking, inspired by the WATCH
workflow \citep{Sechidis2025WATCH} and its evaluation via individualized treatment
effects \citep{Sechidis2025ITE}.

\textbf{Individualized treatment effect proxy.}\\ Central to our implementation is the construction of an individualized treatment
effect proxy $\hat{\phi}_i$ for each patient. We employ the DR-learner
\citep{kennedy2023towards}, which combines cross-fitted propensity score and outcome
models to produce a debiased pseudo-outcome interpretable as an observation of the
patient-level conditional average treatment effect; see
Sechidis et al. 2025a \citep{Sechidis2025WATCH} for implementation details and
Sechidis et al. 2025b \citep{Sechidis2025ITE} for a comprehensive evaluation. In this step, region is included as part of the baseline covariate set used to estimate the conditional treatment effect. Its role in the downstream workflow then depends on the question being addressed. When treatment effects are defined on model-based scales (e.g., hazard ratios or odds ratios), score
residuals of the treatment effect parameter from a fitted regression model
\citep{chen2026comparing} can serve the same role. The workflow structure is
agnostic to this choice: what matters is that $\hat{\phi}$ captures
individual-level deviations from treatment effect homogeneity. Once constructed,
the same $\hat{\phi}$ is used throughout Q1--Q4, providing a unified analytical
backbone across all four questions.

\textbf{Running example for illustration}\\ To illustrate the workflow, we use simulated data from the \texttt{benchtm} R
package~\citep{Sun2024bimj}, which generates synthetic baseline covariates preserving the correlation structure of real clinical trial data. A sample of $n = 500$ patients is generated with $p = 30$ baseline covariates $X_1, X_2, \ldots, X_{30}$ and balanced randomization $P(Z = 1) = P(Z = 0) = 0.5$. The continuous outcome follows $Y \sim N(\mu,\, \sigma = 1)$ with
$\mu = 2.32 \times \{0.5(\mathbb{I}(X_1 = \text{"Y"})+ X_{11})\} + Z\{-0.106 + 0.767\Phi(20(X_{11}-0.5))\}$, where $\Phi(\cdot)$ is the cumulative distribution function (cdf) of the standard normal distribution.
A binary region label is generated via a logistic model in $X_{11}$, with the intercept calibrated so that the marginal prevalence $P(\mathit{Region} = 1) \approx 0.2$ and a slope corresponding to an odds ratio of 10 for the $X_{11}$–region association. By construction, $X_{11}$ is both a treatment effect modifier and regionally imbalanced, making it a region-associated effect modifier ($\mathcal{X}_2$ in Section~\ref{sec:framework}).

To illustrate two cases that may arise in practice, the workflow is applied to the \emph{same data} with $X_{11}$ either included or excluded from the analysis covariate set:
(i)~in the observed case, $X_{11}$ is available and regional heterogeneity is plausibly  explainable by an observed region-associated effect modifier~($\mathcal{X}_2$);
(ii)~in the unobserved case, $X_{11}$ is absent from the analysis covariate set, mimicking a situation where the true effect modifier was not measured,so that it acts as an unobserved factor~($U$) for which region may serve as a proxy.

The resulting region-specific treatment effect estimates from a treatment-by-region interaction model differ
noticeably ($\hat{\delta}_{\mathit{Region}=0}=\text{0.130}$, 95\%
CI [-0.116, 0.376];
$\hat{\delta}_{\mathit{Region}=1}=\text{0.618}$, 95\% CI [0.169,
1.067]), motivating a structured investigation via Q1--Q4.

\textbf{Preprocessing and alignment with the primary analysis}\\
Before proceeding to the analysis, we note that a rigorous
post-hoc investigation requires alignment with the estimand and
endpoint definition from the primary analysis, as well as careful
curation of the baseline covariate set. General guidance on
covariate selection for treatment effect heterogeneity assessments
is provided in the WATCH workflow~\citep{Sechidis2025WATCH}; in
the MRCT setting, covariate selection should additionally be
informed by intrinsic and extrinsic factors identified in
ICH~E17~\citep{ICH2017} as potential sources of regional
differences, as well as prior knowledge of the treatment's
mechanism of action, the disease, and known differences across the
participating regions. A purely data-driven selection that includes
all available variables without clinical rationale risks
identifying spurious associations and complicates interpretation.
Typical preprocessing steps include transformations for skewed
distributions, consolidation of sparse categories, and principled
imputation of missing baseline data; see Sechidis et
al.~\citep{Sechidis2025WATCH} and Baillie et
al.~\citep{baillie2022ten} for detailed guidance. All
preprocessing decisions should be documented to ensure
reproducibility.

\begin{figure*}
\centerline{\includegraphics[width=475pt,height=25pc]{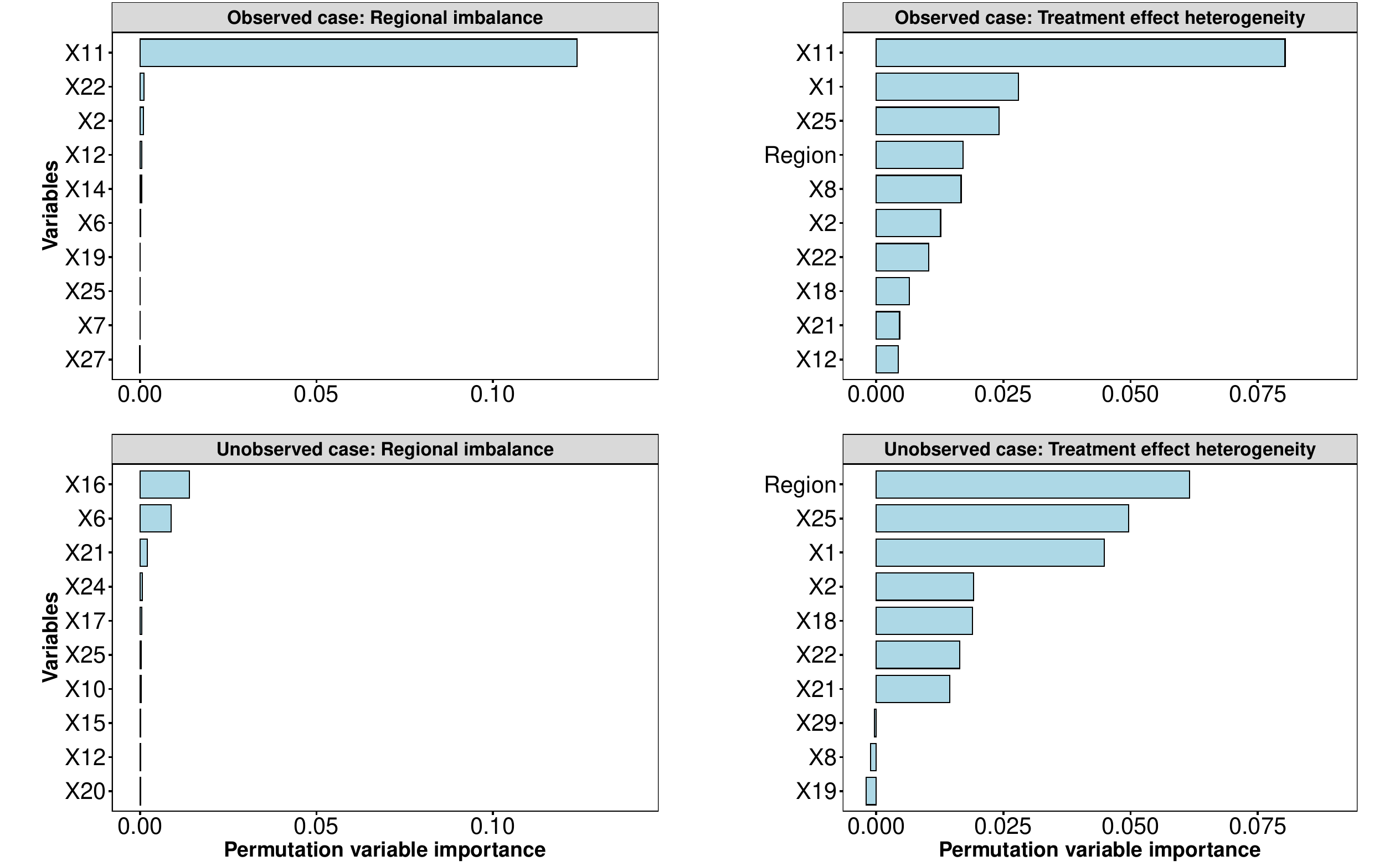}}
\caption{\textbf{Running example: variable importance rankings.} Four panels correspond to Scenario~1 in
Table~\ref{tab:benchtm_sc12}. The upper row displays results under the observed case, the lower row under the unobserved case. For each case, the left panel presents the top covariates associated with regional imbalance (Q2), whereas the right panel shows the top covariates associated with treatment effect heterogeneity (Q3).
Under the observed case, $X_{11}$ ranks first in both Q2 and Q3.
Under the unobserved case, Q2
importance scores are an order of magnitude smaller and no single
covariate dominates, while in Q3 $\mathit{Region}$ rises to the
top, consistent with its proxy role for the excluded effect
modifier. \label{fig:Q2-vi}}
\end{figure*}

\begin{figure*}
\centerline{\includegraphics[width=300pt]{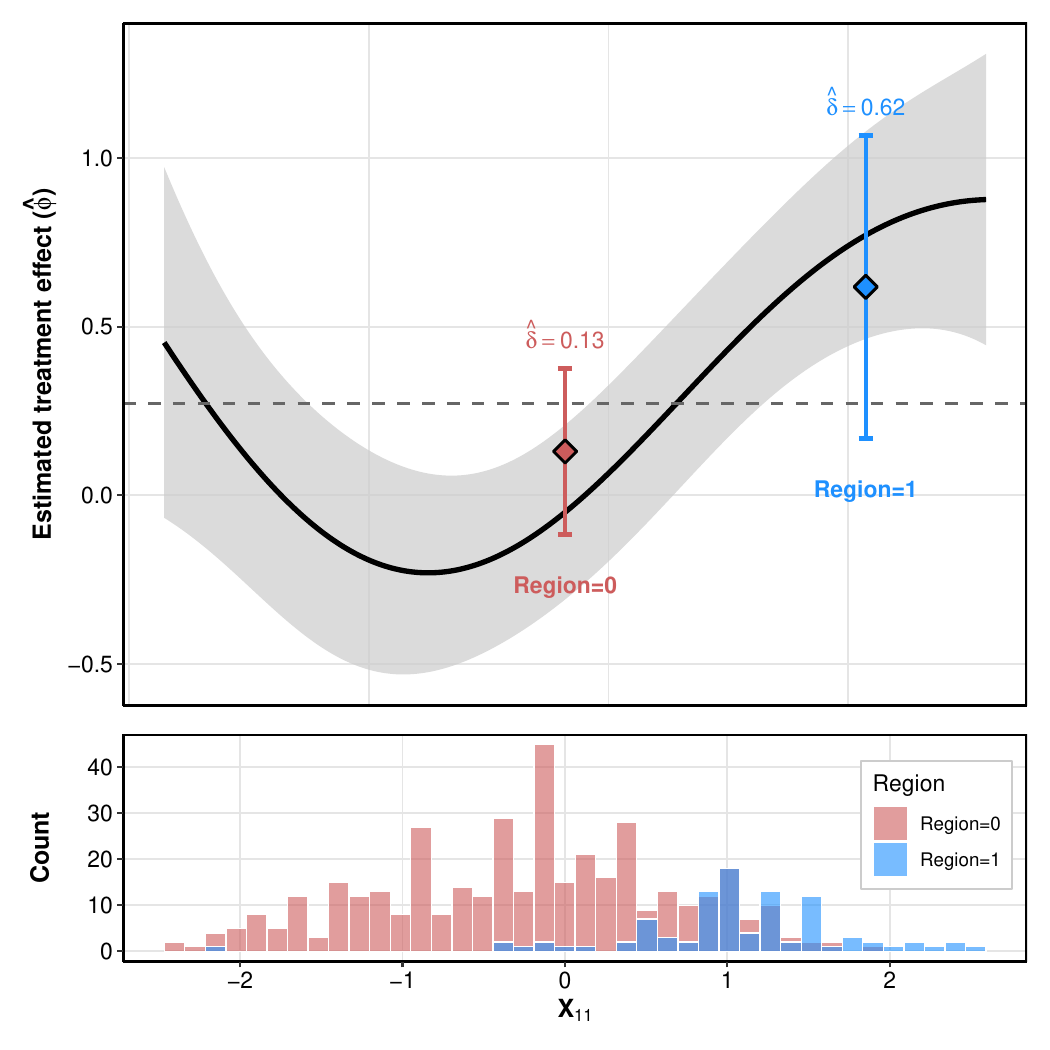}}
\caption{\textbf{Q4 display for the leading regional effect
modifier candidate $X_{11}$.}
\textit{Upper panel:} estimated treatment effect ($\hat{\phi}$)
as a function of $X_{11}$, fitted using a penalized spline via a generalized additive model (GAM; smoothing selected by REML) , with pointwise 95\% confidence band. The horizontal dashed line marks the overall average treatment effect. Diamond markers show region-specific treatment effect estimates from a treatment-by-region interaction model
($\hat{\delta}_{\mathit{Region}=k}$), positioned at the
region-specific median of $X_{11}$, with 95\% confidence
intervals. The smooth curve is displayed from the 11th to the
$(n{-}10)$th ordered value of $X_{11}$ to avoid unreliable
boundary estimates in data-sparse regions. These marginal estimates average over each region's
full covariate distribution and thus need not coincide with the
conditional curve at the median.
\textit{Lower panel:} distribution of $X_{11}$ by region
(red: $\mathit{Region}=0$; blue: $\mathit{Region}=1$),
illustrating the distributional shift that links $X_{11}$ to
regional heterogeneity. The regional difference in treatment
effects is consistent with the distributional separation:
$\mathit{Region}=1$ patients are concentrated in
higher-$X_{11}$ ranges where the estimated treatment effect is
larger.
\label{fig:Q4-display}}
\end{figure*}

\subsection{Q1: Is there evidence of regional heterogeneity in treatment
effects?}\label{sec:Q1}

The first question asks whether the distribution of individual treatment effects
differs across regions. Given the pseudo-outcome $\hat{\phi}$ constructed above,
we test
\begin{equation}\label{eq:H0-RV}
  H_0\colon \hat{\phi} \perp\!\!\!\perp \mathit{Region},
\end{equation}
using a permutation-based conditional inference framework with the maximum
statistic \citep{hothorn2006lego}, and denote the resulting $p$-value as
$p_{\mathrm{RV}}$. A common alternative is to embed a treatment-by-region
interaction term in a parametric outcome model; the pseudo-outcome formulation
avoids the need to specify such a parametric interaction structure and
accommodates nonlinear heterogeneity patterns. As outlined in
Section~\ref{sec:framework}, $p_{\mathrm{RV}}$ is interpreted on a continuous
scale; when $p_{\mathrm{RV}}$ is large, subsequent region-associated effect modifier identification
(Q2--Q3) will be less reliable and findings should be interpreted with greater
caution. Complementary descriptions such as funnel plots  are recommended to contextualize
$p_{\mathrm{RV}}$ and support interpretation.

\textbf{Running example.}\\
Under both cases, $p_{\mathrm{RV}}$ indicates strong evidence against regional
homogeneity ($p_{\mathrm{RV}} = \text{0.0037}$). This concordance is expected: Q1 assesses the presence of regional heterogeneity without reference to any specific covariate, so $p_{\mathrm{RV}}$ is largely determined by the association between $\hat{\phi}$ and $\mathit{Region}$ rather than by which covariates enter the outcome model. The two cases diverge only in subsequent questions, where the ability
to explain the observed pattern depends on the available covariate
information. Following the decision roadmap (Table~\ref{tab:roadmap}), strong
evidence at Node~1 motivates proceeding to Q2 and Q3.

\subsection{Q2: Which baseline covariates are unevenly distributed across regions?}\label{sec:Q2}

Given evidence of regional heterogeneity from Q1, Q2 assesses whether observed
baseline covariates are imbalanced across regions. We first summarize the overall
evidence of regional imbalance by performing a permutation-based global
independence test of $\mathit{Region} \perp\!\!\!\perp \mathbf{X}$, yielding
$p_{\mathrm{RI}}$ as a summary measure. As in Q1, $p_{\mathrm{RI}}$ is
interpreted on a continuous scale; it can be very small when regions correspond
to populations with markedly different demographic or clinical baseline characteristics.

To identify which covariates best distinguish regions, we model
$\mathit{Region} \sim \mathbf{X}$ using a conditional random forest (CRF) and
extract permutation-based variable importance (VI) scores. This ranking is based on a multivariate model accounting for correlations among covariates, in contrast to univariate
screening approaches (e.g., per-variable standardized mean differences).
Covariates with high importance correspond to candidates in
$\mathcal{X}_1 \cup \mathcal{X}_2$ and are carried forward to the overlap
assessment in Q4.

\textbf{Running example.}\\
Under both cases, $p_{\mathrm{RI}}$ is very small ($p < 0.0001$),
confirming that the regional populations differ on measured baseline covariates.
Under observed case, the CRF ranking highlights $X_{11}$ as the dominant predictor
of $\mathit{Region}$, consistent with the data-generating mechanism. Under the unobserved case, where $X_{11}$ is excluded, not only is the ranking diffuse, but the absolute variable importance scores are markedly smaller, indicating that the remaining covariates carry little information for distinguishing regions (see Figure~\ref{fig:Q2-vi}).

The simulation study in Section~\ref{sec:simulation} further examines settings
where correlated covariates may serve as proxies for the missing variable.
Following the decision roadmap (Table~\ref{tab:roadmap}), Node~2 records the
ranked list and carries it forward to the overlap assessment.

\subsection{Q3: Which baseline covariates modify the treatment effect?}
\label{sec:Q3}

Q3 mirrors the structure of Q2 but targets treatment effect modification rather
than regional population structure. This question is well known to be very challenging to answer reliably given the sample sizes available in a single trial~\citep{kent2020a}. A global independence test of
$\hat{\phi} \perp\!\!\!\perp \mathbf{X}$ yields $p_{\mathrm{TEH}}$, which
quantifies the overall evidence against treatment effect homogeneity across all
baseline covariates. This test differs from the Q1 test: $p_{\mathrm{RV}}$
assesses whether $\hat{\phi}$ varies across regions (a single categorical
variable), whereas $p_{\mathrm{TEH}}$ assesses whether $\hat{\phi}$ depends on
any baseline covariate. Using the pseudo-outcome $\hat{\phi}$ from Q1, we then
model
$$\hat{\phi} \sim \mathbf{X} + \mathit{Region},$$
using a conditional random forest and extract permutation-based VI scores to
rank covariates by the degree to which they modify the treatment effect.
$\mathit{Region}$ is included as a covariate so that its relative importance
can be compared directly with that of measured baseline variables: a high VI
for $\mathit{Region}$ suggests that the regional grouping captures
heterogeneity not explained by the observed covariates, consistent with the
surrogate role of $\mathit{Region}$ described in Section~\ref{sec:conceptual}.
Covariates with high VI correspond to candidates in
$\mathcal{X}_2 \cup \mathcal{X}_3$ and are carried forward to the overlap
assessment in Q4.

\textbf{Running example.}\\
Under observed case, $p_{\mathrm{TEH}}$ indicates modest evidence against homogeneity
($p_{\mathrm{TEH}} = \text{0.184}$). The VI ranking
(Figure~\ref{fig:Q2-vi}) places $X_{11}$ first, followed by $X_1$,
$X_{25}$, and $\mathit{Region}$ at rank~4. 
Under unobserved case, $p_{\mathrm{TEH}} = \text{0.232}$; notably,
$\mathit{Region}$ rises to rank~1, illustrating 
the surrogate mechanism from the conceptual framework
(Figure~\ref{fig1}): when the true modifier is absent, $\mathit{Region}$
absorbs part of its explanatory role via the pathway
$\mathit{Region}$--$U$--$TE$. Following the decision roadmap
(Table~\ref{tab:roadmap}), Node~3 records the ranked list of treatment effect modifiers
and carries it forward to the overlap assessment.

\subsection{Q4: How do treatment effects vary along key effect modifiers across regions?}\label{sec:Q4}

The final question synthesizes the outputs of Q2 and Q3 by 
examining the overlap between region-associated covariates 
and treatment effect modifiers. Covariates appearing in both 
ranked lists are candidates for $\mathcal{X}_2$ (regional 
effect modifiers) and are prioritized for descriptive displays. 
When the overlap is empty, the observed regional heterogeneity 
cannot be readily attributed to measured covariates, and 
$\mathit{Region}$ may be acting as a surrogate for unmeasured 
factors (Table~\ref{tab:roadmap}, Node~4).

For each prioritized covariate $X_j$, we recommend a combined 
display with two panels. The upper panel shows a smooth 
estimate of $\hat{\phi}$ versus $X_j$ fitted to the full 
dataset, with region-specific location markers (e.g., medians) 
indicating where each region's patients fall along the 
covariate axis. The lower panel shows the distribution of 
$X_j$ separately by region (e.g., overlaid densities or 
side-by-side histograms). This design separates two distinct 
sources of regional differences: (a)~non-overlapping covariate 
distributions across regions (population structure) and 
(b)~differing treatment effects within shared covariate ranges 
(heterogeneous response). When regional sample sizes are 
sufficient, region-specific summaries such as binned means 
with confidence intervals or region-separate fitted curves may be overlaid to assess 
whether regional treatment effects deviate from the overall 
pattern; for small regions such summaries can be unstable and 
are not recommended.

Per-arm outcome plots displaying mean outcomes on treatment 
and control arms separately as a function of $X_j$, can 
help clarify whether regional differences are driven by the 
treatment arm, the control arm, or both. This distinction 
supports interpretation of the mechanism underlying the 
observed heterogeneity. When a flagged regional signal involves 
many countries or numerous small regions, country-level 
descriptive plots (e.g., radial or funnel plots) can help 
contextualize outlying effects relative to their uncertainty.

\textbf{Running example.}\\
Figure~\ref{fig:Q4-display} presents the Q4 display for
$X_{11}$, the leading candidate from the Q2/Q3 overlap under
the observed case. The upper panel shows that the treatment
effect increases with $X_{11}$; the lower panel reveals that
$\mathit{Region}=1$ patients are concentrated in the
higher-$X_{11}$ range where the treatment effect is larger,
while $\mathit{Region}=0$ patients are spread across lower
values. This pattern is consistent with the interpretation
that the observed regional heterogeneity in treatment effects
can be attributed to the regional imbalance in $X_{11}$ rather
than to a genuinely different treatment response. Under the
unobserved case, where $X_{11}$ is excluded from the analysis
set, no candidate emerges from the Q2/Q3 overlap and the
display cannot be constructed, consistent with the terminal
conclusion of unexplained heterogeneity (T4 in
Table~\ref{tab:roadmap}).

\subsection{Practical considerations}\label{sec:practical}

\textbf{Multidisciplinary assessment.}\\
The outputs from Q1--Q4 are exploratory in nature and should be 
reviewed by a multidisciplinary team encompassing statistical, 
clinical, and regulatory expertise. Such review serves to evaluate 
whether identified candidates represent plausible intrinsic or 
extrinsic factors of regional heterogeneity, integrating a priori 
evidence from historical clinical data, external literature, and 
scientific understanding of the disease and treatment mechanism. 
Equally important, multidisciplinary assessment guards against 
over-interpretation of data-driven findings from a single trial 
and ensures that conclusions are proportionate to the strength 
and consistency of the observed signals. 

\textbf{Optional extensions.}\\
When some regions contribute only limited sample sizes, 
region-specific treatment effect estimates can be stabilized 
through partial pooling or shrinkage approaches that borrow 
strength across regions under explicit modeling assumptions 
\citep{Quan2013EmpiricalShrinkage, Quan2014MRCTDesignConsistency}; such methods may be 
particularly relevant when local regulatory assessment focuses 
on smaller regions. A natural further question is to what extent 
the identified region-associated effect modifiers quantitatively account 
for the observed heterogeneity. Approaches inspired by mediation 
decomposition or covariate standardization can in principle 
address this question within the pseudo-outcome framework used 
here; however, their properties in the MRCT setting warrant 
careful investigation and are deferred to future work.

\section{Simulation study}\label{sec:simulation}

We conducted a simulation study to evaluate the operating characteristics of the proposed methods for explaining region-associated treatment effect heterogeneity in MRCTs. We structured the simulation according to the ADEMP framework (Aims, Data-generating mechanisms, Targets, Methods, and Performance measures).

\subsection{Aims}
We evaluate the methods across realistic and challenging MRCT scenarios motivated by the conceptual framework in 
Section~\ref{sec:conceptual}, including (i) no region-associated treatment effect heterogeneity, (ii) regional heterogeneity driven by observed region-associated effect modifiers, and (iii) regional heterogeneity driven by unobserved variables with correlated observed baseline variables (proxy variables).  We assess both whether the  models yield appropriate conclusions under known ground truth, and how its output patterns, such as global evidence summaries and covariate rankings,  vary across these cases to guide interpretation by MRCT study teams. 

\subsection{Data-generating mechanisms}

\textbf{Baseline covariates and outcome model.}
We generated baseline covariates $\mathbf{X}=(X_1,\ldots,X_{30})$ using the \texttt{benchtm} R package \citep{benchtm}, which mimics multivariate covariate distributions observed in clinical trial data and has been used previously to benchmark methods for treatment effect heterogeneity in realistic settings. A sample size of $n = 500$ is used 
with balanced randomization $\Pr(Z = 1) = 0.5$. We focus on continuous outcomes and adopt \texttt{benchtm} Scenarios~1--2, where TEH is driven by a single continuous predictive covariate: $X_{11}$ in Scenario~1 and $X_{14}$ in Scenario~2 \citep{Sechidis2025ITE,Sun2024bimj}. 

Outcomes are generated from the general form
\[
Y = f_{\text{prog}}(\mathbf{X}) + Z\big(\beta_0 + \beta_1 f_{\text{pred}}(\mathbf{X})\big) + \varepsilon,\qquad \varepsilon\sim N(0,1),
\]
where $Z\in\{0,1\}$ denotes randomized treatment assignment with $\Pr(Z=1)=0.5$, $f_{\text{prog}}(\mathbf{X})$ captures prognostic structure, and $f_{\text{pred}}(\mathbf{X})$ captures predictive structure that modifies the treatment effect \citep{Sechidis2025ITE,Sun2024bimj}. 

\begin{table}[t]
\centering
\begin{tabular}{ll}
\hline
No. & $f(\mathbf{X},Z)$ \\
\hline
Scenario 1 & $s\cdot\{0.5\,\mathbb{I}(X_1=Y)+X_{11}\}+Z\{\beta_0+\beta_1\Phi(20(X_{11}-0.5))\}$ \\
Scenario 2 & $s\cdot\{X_{14}-\mathbb{I}(X_8=N)\}+Z\{\beta_0+\beta_1 X_{14}\}$ \\
\hline
\end{tabular}
\begin{tablenotes}
\footnotesize
\caption{Outcome-generating models for the two \texttt{benchtm} scenarios used in this paper (continuous endpoint).}
\label{tab:benchtm_sc12}
\item \textit{Note:} $\mathbb{I}(\cdot)$ is the indicator function, $\Phi(\cdot)$ is the cumulative distribution function (cdf) of the standard normal distribution, and $s$ is a scaling factor chosen (scenario-specifically) so that a prespecified $R^2$ is achieved on the control arm ($Z=0$). See Sun et al. for details on calibration of $s$.
\end{tablenotes}
\end{table}

In Scenario~1, the predictive component is a smooth step-like function of $X_{11}$ via a probit function, whereas in Scenario~2 it is linear in $X_{14}$ (Table~\ref{tab:benchtm_sc12}). The parameter $\beta_1$ governs the strength and hence detectability of TEH, while $\beta_0$ controls the overall treatment effect level. Following the benchmarking rationale in \texttt{benchtm}, $\beta_1^\ast$ is defined as the interaction magnitude yielding a power of 80\% for detecting effect modification under the true data-generating model, and vary TEH strength via the ratio $\beta_1/\beta_1^\ast$ \citep{Sun2024bimj}. In our simulations we vary the treatment interaction strength across a prespecified grid of interaction strengths. Details on calibration of model parameters and data pre-process were described previously \citep{Sun2024bimj}.

\textbf{Region generation.}
A binary regional indicator $Region\in\{0,1\}$ with target prevalence $\Pr(Region=1)=0.2$ is generated using a logistic model based on the treatment effect modifier that drives TEH:
\[
\text{logit}\{\Pr(Region=1\mid X_{\text{pred}})\}=\alpha_0+\alpha_1 X_{\text{pred}},
\]
where $X_{\text{pred}}=X_{11}$ in Scenario~1 and $X_{\text{pred}}=X_{14}$ in Scenario~2.  The slope 
$\alpha_1$ is chosen to achieve a prespecified odds ratio 
(OR) for the $X_{\text{pred}}$--region association, and the 
intercept $\alpha_0$ is set to maintain the target regional 
prevalence. By construction, $X_{\text{pred}}$ is both a 
treatment effect modifier and regionally imbalanced, making 
it a region-associated effect modifier ($\mathcal{X}_2$). Because numeric covariates are scaled to $[0,1]$, the region-imbalance parameterization via odds ratios yields comparable effect sizes across scenarios and across candidate predictors. 

\textbf{Observed versus unobserved cases with proxy covariates.}
Each simulated dataset is analyzed under two cases that 
differ only in the covariate set provided. In the observed case, the region-associated effect modifier $X_{\text{pred}}$ (i.e., $X_{11}$ in Scenario~1 or $X_{14}$ in Scenario~2) is included in the analysis covariate vector $\mathbf{X}$. In the unobserved case, $X_{\text{pred}}$ is excluded from the analysis covariate vector, while the data-generating mechanism for $(Y,Region)$ remains unchanged. In this unobserved case, other observed covariates that are correlated with the omitted $X_{\text{pred}}$ may act as proxy covariates and partially inherit its association with $Region$ and/or with the treatment-effect signal. Here, proxy covariates refers to observed baseline covariates correlated with the omitted factor $X_{\text{pred}}$.  The proxy structure differs markedly between scenarios. In Scenario~1, $X_{11}$ is only weakly correlated with other
covariates (mean $|\hat{\rho}| \leq 0.17$ across replications),
implying no usable proxy; in Scenario~2, $X_{14}$ is strongly
correlated with $X_{12}$ (mean $|\hat{\rho}| = 0.86$) and
moderately with $X_9$ (mean $|\hat{\rho}| = 0.41$), which supports
interpretation of proxy-driven ranking patterns; see
Table~S1 in the Supplementary Material for details. Because the
covariate set includes both continuous and categorical variables,
pairwise associations are estimated via latent Gaussian correlations
using the \texttt{latentcor} R
package~\citep{latentcor_reference}.

The simulation design crosses five interaction strengths
$\beta_1/\beta_1^\ast \in \{0, 0.5, 1, 1.5, 2\}$ with five
regional imbalance levels $\mathrm{OR} \in \{1, 1.5, 2, 5, 10\}$
and two predictive-function scenarios (Scenarios~1 and~2), yielding
$5 \times 5 \times 2 = 50$ data-generating configurations; here
$\beta_1/\beta_1^\ast = 0$ corresponds to no treatment effect
heterogeneity and $\mathrm{OR} = 1$ to no association between
$\mathit{Region}$ and $X_{\text{pred}}$. Each configuration is
analyzed under both the observed and unobserved cases, which share
the same data-generating mechanism but differ in the covariate set
available to the analyst (see the running example in Section~\ref{methods-example}), giving
$50 \times 2 = 100$ configuration--case combinations in total. For
each combination, $R = 500$ independent datasets are generated and
the results are summarized across replications.

\subsection{Methods of analysis}\label{sec:sim-method}
Each simulated dataset is analyzed using the methods described in Section~\ref{methods-example} (Steps~1--3) with all tuning and implementation choices fixed across scenarios and analysis-input cases. In brief, we (i) construct the doubly robust pseudo-observation $\hat{\phi}$ and obtain a permutation-based global independence $p$-value for $\hat{\phi}$ versus $Region$ ($p_{\mathrm{RV}}$); (ii) obtain a regional-covariate ranking from a CRF model for $Region\sim\mathbf{X}$ together with a global $p$-value summarizing evidence for regional imbalance ($p_{\mathrm{RI}}$); and (iii) obtain an effect-modifier ranking from a CRF model for $\hat{\phi}$ using the analysis covariate set (and $Region$ when included), optionally recording a global TEH evidence summary ($p_{\mathrm{TEH}}$). Let $\mathcal{T}_{\mathrm{Reg}}(K)$ and 
$\mathcal{T}_{\mathrm{EM}}(K)$ denote the top-$K$ covariates 
from the Q2 and Q3 rankings, respectively. Candidate region-associated effect modifiers are identified via the 
intersection 
$\mathcal{T}_{\cap} = 
\mathcal{T}_{\mathrm{Reg}}(5) \cap 
\mathcal{T}_{\mathrm{EM}}(5)$, 
which contains between $0$ and $5$ covariates.

\subsection{Performance measures}
\label{sec:sim-performance}

\textbf{Objective 1: Calibration and sensitivity of global 
evidence summaries.}\\
We assess the three $p$-values produced by the proposed method 
($p_{\mathrm{RV}}$ from Q1, $p_{\mathrm{RI}}$ from Q2, 
$p_{\mathrm{TEH}}$ from Q3). Under settings with no 
corresponding signal, each $p$-value should be approximately 
Uniform$(0,1)$; we assess this using empirical cumulative 
distribution functions (ECDFs) against the uniform diagonal. 
Under settings with signal, smaller $p$-values indicate 
higher sensitivity; we summarize this using the median 
surprise value, $\mathrm{median}\{-\log_2(p)\}$, where 
larger values correspond to stronger evidence against 
homogeneity~\citep{cole_2020}.

\textbf{Objective 2: Recovery of regional effect 
modifiers.}
Using the notation from Section~\ref{sec:sim-method}, we 
evaluate identification of region-associated effect modifiers at two 
levels of stringency: top-1 hit rates (requiring the target 
covariate to be ranked first in $\mathcal{T}_{\mathrm{Reg}}$ or $\mathcal{T}_{\mathrm{EM}}$) and overlap recovery 
(membership in $\mathcal{T}_{\cap}(5)$). What constitutes the 
target depends on the evaluation case: under case~(i), no 
covariate is a true region-associated effect modifier, and an unbiased 
method should select each of the  covariates with 
approximately equal probability; under case~(ii), the 
target is the observed region-associated effect modifier $X_{\text{pred}}$; 
under case~(iii), truth-based hit rates for the excluded 
$X_{\text{pred}}$ are not defined, and we instead evaluate 
whether proxy covariates or $\mathit{Region}$ itself emerge 
in the rankings and overlap. To complement hit-rate summaries, 
we report per-variable selection probability profiles across 
both rankings and their intersection, which reveal how 
individual covariates are prioritized as signal strength and 
regional imbalance vary.

\subsection{Simulation results}
\label{sec:sim-results}

\subsubsection{Objective 1: Calibration and sensitivity of global evidence summaries}
\label{sec:sim-global}

The three global $p$-values produced by the proposed methods, $p_{\mathrm{RV}}$ (Q1), 
$p_{\mathrm{RI}}$ (Q2), and $p_{\mathrm{TEH}}$ (Q3), provide global evidence summaries that guide the analyst through the decision nodes of the 
roadmap (Table~\ref{tab:roadmap}). We first assess their calibration under 
null-like settings, then examine their sensitivity as signal strength increases.

\begin{figure}[t]
  \centering
  \includegraphics[width=\textwidth]{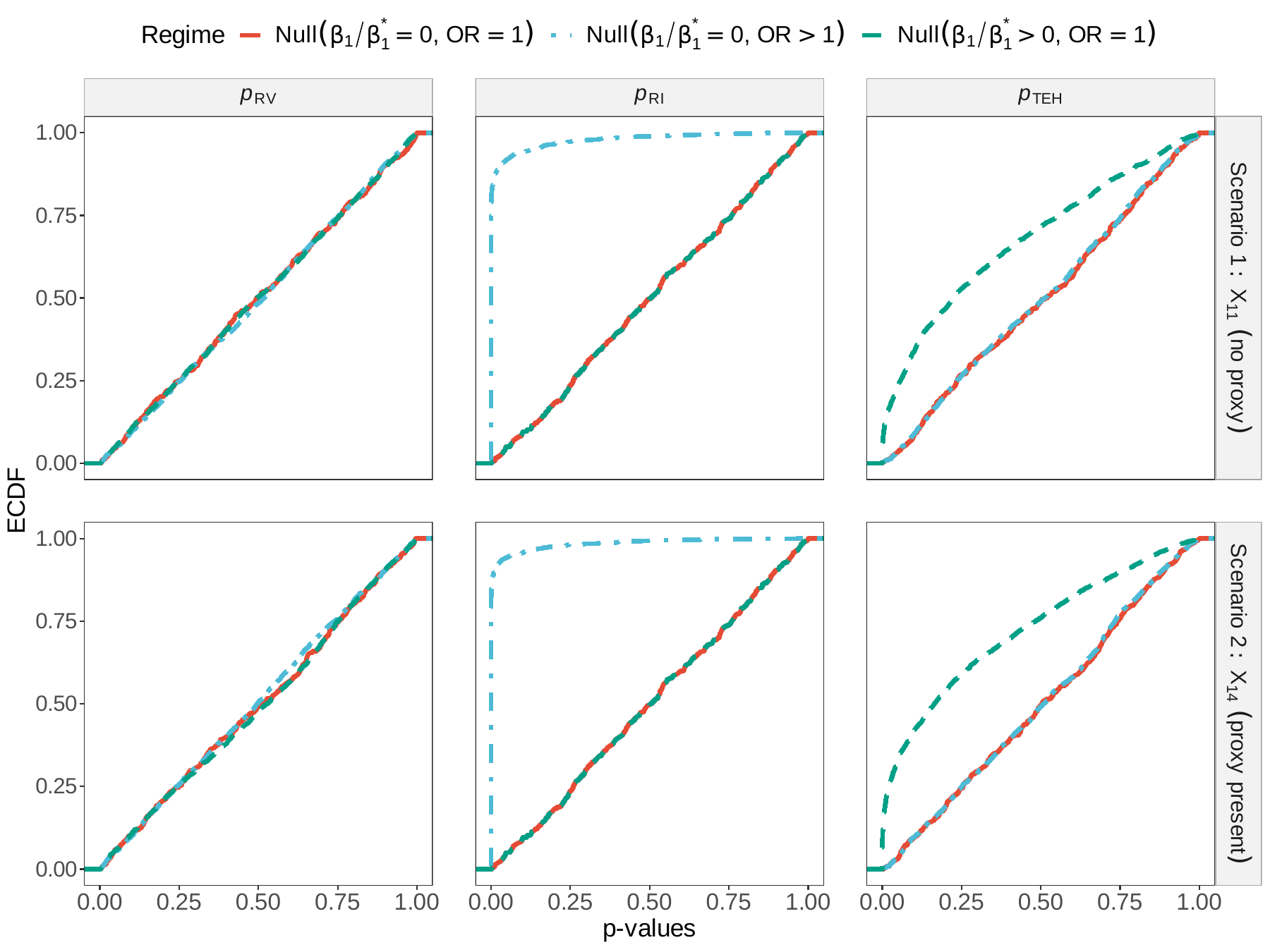}
  \caption{\textbf{Calibration of global evidence summaries under null-like settings.}
  Empirical CDFs of calibrated $p$-values for regional-variability evidence ($p_{\mathrm{RV}}$), regional-imbalance evidence ($p_{\mathrm{RI}}$), and TEH evidence ($p_{\mathrm{TEH}}$) under three null-like settings: $\beta_1/\beta_1^{*}=0$, OR$=1$ (complete homogeneity); $\beta_1/\beta_1^{*}=0$, OR$>1$ (imbalance-only); and $\beta_1/\beta_1^{*}>0$, OR$=1$ (TEH-only). Rows correspond to \texttt{benchtm} Scenarios~1--2 and panels are shown for both observed and unobserved analysis-input cases. The diagonal line indicates the Uniform$(0,1)$ reference.}
  \label{fig:sim-ecdf}
\end{figure}

\textbf{Calibration under null-like settings.}  \\
Figure~\ref{fig:sim-ecdf} displays the calibration of the 
three global $p$-values under null-like settings. Results are shown for the observed 
analysis-input case.
Under complete homogeneity ($\beta_1/\beta_1^{*}=0$, OR$=1$), all 
three summaries were close to Uniform$(0,1)$, confirming 
appropriate type-I error behavior. Under regional imbalance 
without TEH ($\beta_1/\beta_1^{*}=0$, OR$>1$), only 
$p_{\mathrm{RI}}$ (Q2) deviated toward small values, while 
$p_{\mathrm{RV}}$ (Q1) and $p_{\mathrm{TEH}}$ (Q3) remained 
calibrated. Conversely, under TEH without regional imbalance 
($\beta_1/\beta_1^{*}>0$, OR$=1$), only $p_{\mathrm{TEH}}$ 
deviated. These 
patterns were consistent across both predictive-function scenarios (1 and~2) 
and both analysis-input cases (observed and unobserved; Figure~S1). Taken together, the results confirm that each summary responds to 
a distinct component of the workflow's evidence chain: $p_{\mathrm{RV}}$ to 
region-associated heterogeneity, $p_{\mathrm{RI}}$ to covariate imbalance 
across regions, and $p_{\mathrm{TEH}}$ to treatment effect modification by 
baseline covariates. The well-calibrated behavior of $p_{\mathrm{RV}}$ under 
these null-like settings ensures that the method does not spuriously signal 
regional heterogeneity, correctly supporting the conclusion of regional 
consistency (T1 in Table~\ref{tab:roadmap}).

\begin{figure}[t]
  \centering
  \includegraphics[width=\textwidth]{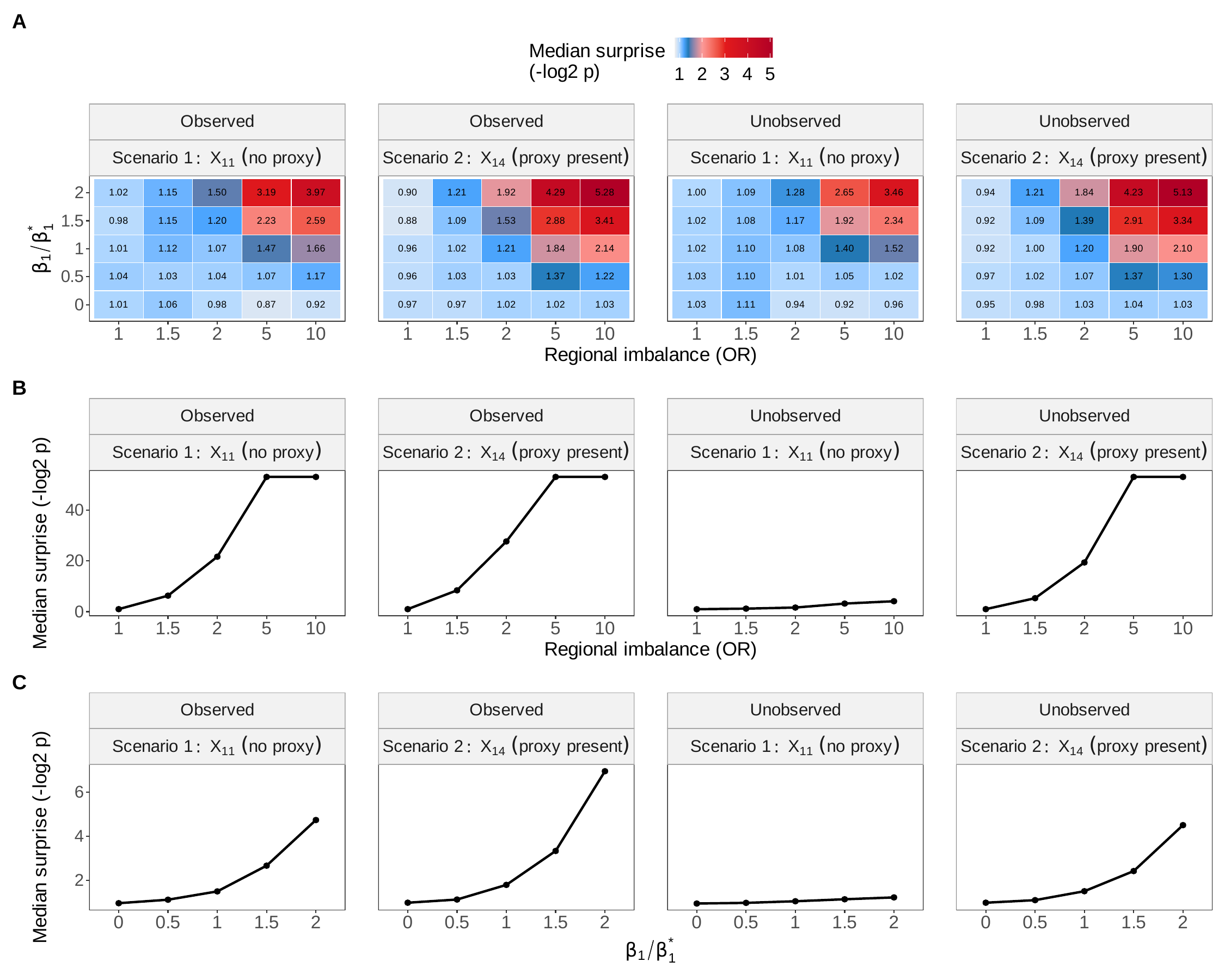}
  \caption{\textbf{Sensitivity of global evidence summaries (median surprise).}
  Median surprise values for $p_{\mathrm{RV}}$ (Panel~A: regional-variability 
  evidence; heatmap over OR and $\beta_1/\beta_1^{*}$), $p_{\mathrm{RI}}$ 
  (Panel~B: regional-imbalance evidence; pooled over $\beta_1/\beta_1^{*}$), 
  and $p_{\mathrm{TEH}}$ (Panel~C: TEH evidence; pooled over OR), shown 
  separately for observed and unobserved analysis-input cases and for 
  predictive-function Scenarios~1--2. Larger values correspond to stronger 
  evidence against homogeneity.}
  \label{fig:sim-surprise}
\end{figure}

\textbf{Sensitivity across the factorial grid.}\\
Figure~\ref{fig:sim-surprise} reports sensitivity via the median surprise value 
across the factorial grid of $\beta_1/\beta_1^{*}$ and OR. 

The Q1 summary $p_{\mathrm{RV}}$, which provides global evidence against 
regional homogeneity, exhibited increasing evidence with both stronger 
interaction and stronger regional imbalance 
(Figure~\ref{fig:sim-surprise}A). Notably, $p_{\mathrm{RV}}$ was virtually identical under both analysis cases, as it depends only on the association between $\hat{\phi}$ and $\mathit{Region}$, not on which covariates are available for the outcome model (as discussed in
Section~\ref{sec:Q1}). The 
Q2 summary $p_{\mathrm{RI}}$, which provides global evidence for regional 
covariate imbalance, depended primarily on OR 
(Figure~\ref{fig:sim-surprise}B), while the Q3 summary $p_{\mathrm{TEH}}$, 
which provides global evidence for treatment effect modification, depended 
primarily on $\beta_1/\beta_1^{*}$ 
(Figure~\ref{fig:sim-surprise}C).

Under the unobserved case, the behavior of 
$p_{\mathrm{RI}}$ and $p_{\mathrm{TEH}}$ diverged across 
scenarios. In Scenario~1 (no usable proxy), both summaries 
were essentially flat, indicating that neither regional 
imbalance nor treatment effect modification could be detected 
from the remaining covariates. In Scenario~2, where the 
strong proxy $X_{12}$ was available, $p_{\mathrm{RI}}$ 
retained sensitivity comparable to the observed case, 
whereas $p_{\mathrm{TEH}}$ showed attenuated but 
appreciable sensitivity. 

These sensitivity patterns have direct implications for the roadmap: when 
$p_{\mathrm{RV}}$ signals regional heterogeneity but the true effect modifier 
is unobserved, the analyst's trajectory depends on the available proxy 
structure. In Scenario~1, the flat sensitivity of both $p_{\mathrm{RI}}$ and 
$p_{\mathrm{TEH}}$ would lead toward T4 (unexplained heterogeneity), whereas 
in Scenario~2, retained sensitivity through proxy covariates supports 
progression toward T3 (partial explanation via observed surrogates). These 
patterns foreshadow the identification results in Objective~2: recovery of 
region-associated effect modifiers via the Q2/Q3 overlap (Q4) requires that both the 
regional-imbalance and effect-modification signals be detectable from the 
available covariates.

\subsubsection{Objective 2: Identification of region-associated effect modifiers}
\label{sec:sim-identification}

\begin{figure}[t]
  \centering
  \includegraphics[width=\textwidth]{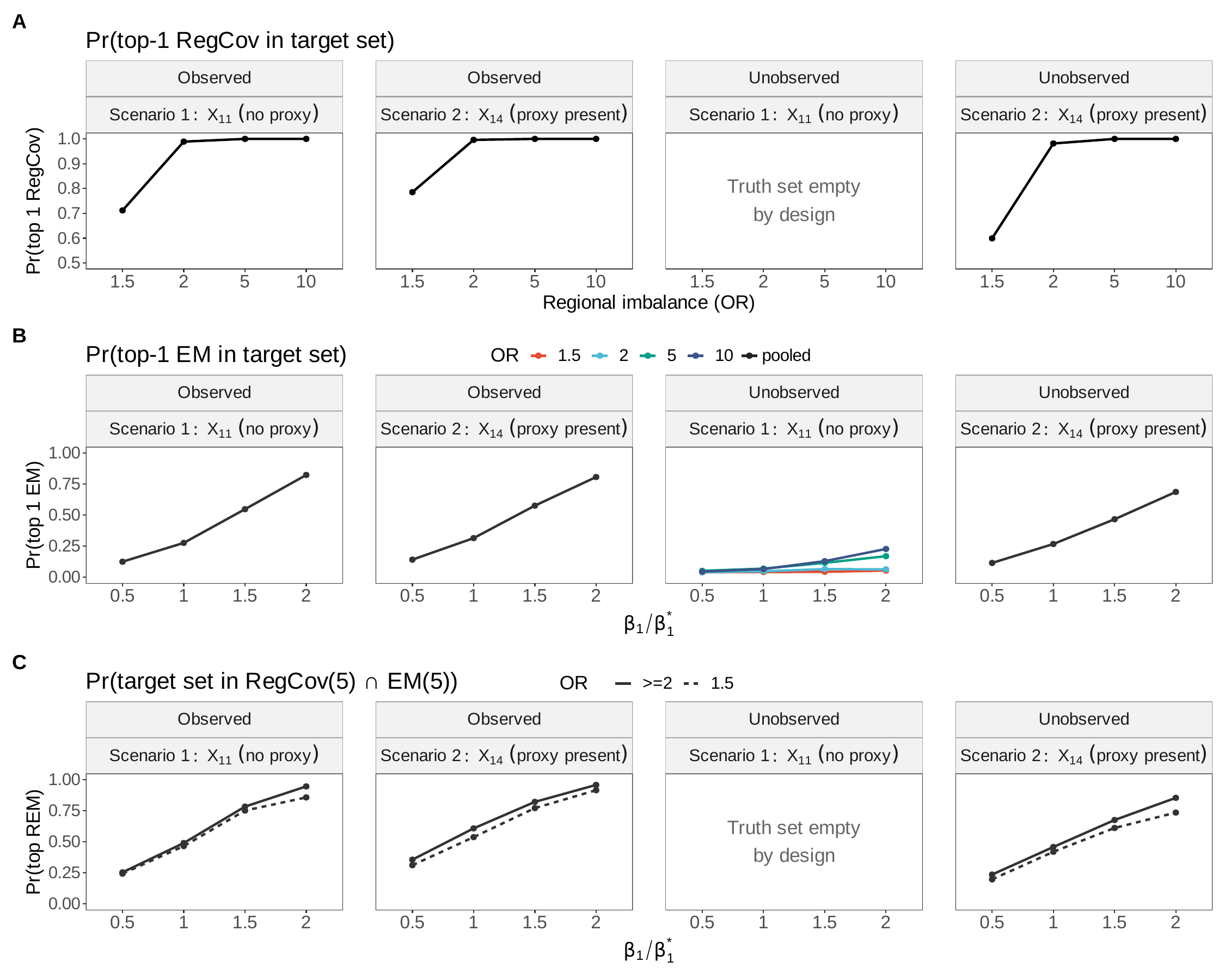}
  \caption{\textbf{Identification performance: top-ranked covariates
and overlap recovery.}
(A)~Probability that the top-1 covariate in the Q2
regional-covariate ranking is in the target, plotted against OR.
(B)~Probability that the top-1 covariate in the Q3
effect-modifier ranking is in the target, plotted against
$\beta_1/\beta_1^{*}$. Black lines: pooled over OR; colored
lines: stratified by OR (shown where the ranking depends on
regional imbalance).
(C)~Probability that the target appears in the overlap set
$\mathcal{T}_{\mathrm{Reg}}(5)\cap\mathcal{T}_{\mathrm{EM}}(5)$,
plotted against $\beta_1/\beta_1^{*}$; solid lines:
OR$\,{\geq}\,2$, dashed lines: OR$\,{=}\,1.5$. In the observed
case, the target is $X_{\text{pred}}$; in the unobserved case,
the target is the set of proxy covariates (joined by
$\mathit{Region}$ in Q3). Results based on 500 replications per
setting.
\label{fig:sim-hit}}
\end{figure}

\begin{figure}[t]
  \centering
  \includegraphics[width=\textwidth]{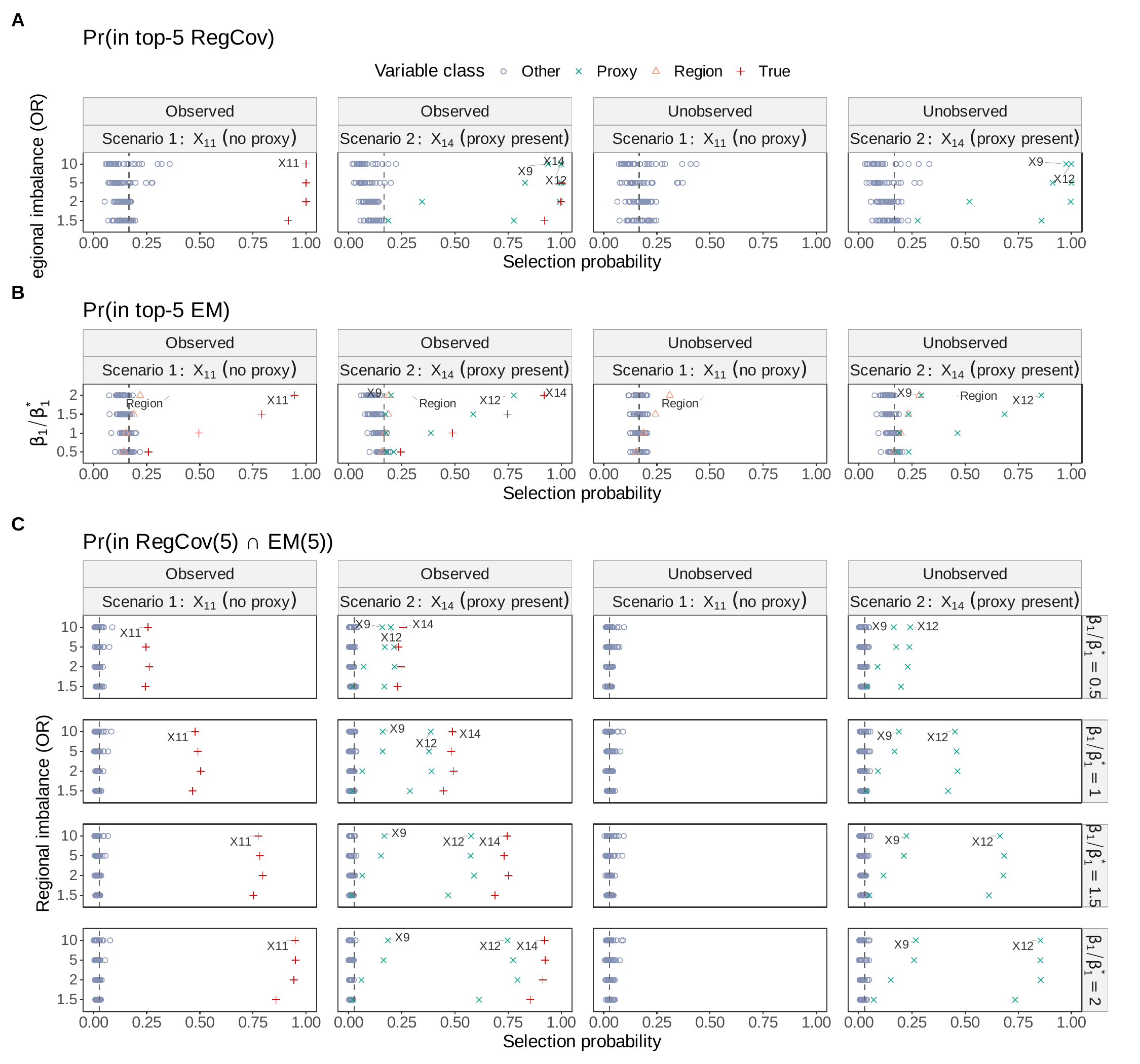}
  \caption{\textbf{Per-variable selection probability profiles (Top-5).}
  Per-variable probabilities of appearing in
  (A)~the top-5 Q2 regional-covariate set $\mathcal{T}_{\mathrm{Reg}}(5)$,
  (B)~the top-5 Q3 effect-modifier set $\mathcal{T}_{\mathrm{EM}}(5)$, and
  (C)~the overlap set
  $\mathcal{T}_{\mathrm{Reg}}(5)\cap\mathcal{T}_{\mathrm{EM}}(5)$.
  Panel~A is stratified by OR (rows); Panel~B by $\beta_1/\beta_1^{*}$
  (rows); Panel~C by both OR (rows) and $\beta_1/\beta_1^{*}$ (column
  facets). All panels are shown for observed and unobserved analysis-input
  cases in Scenarios~1--2. Marker shape and color indicate variable class:
  true effect modifier~(\texttt{+}, red), proxy
  covariates~(\texttt{$\times$}, green),
  $\mathit{Region}$~(\texttt{$\diamond$}, salmon), and other
  covariates~(\texttt{$\circ$}, grey). Note that $\mathit{Region}$ appears
  only in Panels~B and~C, as it serves as the response variable in the Q2
  ranking and is therefore not a candidate in Panel~A.}
  \label{fig:sim-profiles}
\end{figure}

The overlap of the Q2 regional-covariate ranking and the Q3 effect-modifier
ranking forms the basis of Q4, which identifies candidate regional effect
modifiers (Node~4 in Table~\ref{tab:roadmap}). We evaluate identification performance at two levels of resolution. Figure~\ref{fig:sim-hit} provides aggregate summaries: top-1 hit probabilities for the Q2 and Q3 component rankings (the most demanding criterion for individual rankings) alongside recovery in the pre-specified top-5 overlap $\mathcal{T}_{\mathrm{Reg}}(5) \cap \mathcal{T}_{\mathrm{EM}}(5)$. Figure~\ref{fig:sim-profiles} then disaggregates the results to the individual covariate level, showing per-variable top-5 selection probabilities for each component ranking and the overlap. The wider top-5 window in this second figure reveals the behavior of secondary covariates, such as proxy variables and $\mathit{Region}$, that are relevant for interpreting the overlap but may not reach top-1 prominence.

What constitutes the target depends on the evaluation case (see
Section~\ref{sec:sim-performance}): under case~(i), no covariate is a true
region-associated effect modifier, and an unbiased method should select each covariate
with approximately equal probability; under case~(ii), the target is the
observed effect modifier $X_{\text{pred}}$; under case~(iii), truth-based hit
rates for the excluded $X_{\text{pred}}$ are not defined, and we instead
evaluate whether proxy covariates or $\mathit{Region}$ itself emerge in the
rankings. Note that $\mathit{Region}$ is always included as a candidate
covariate in the Q3 effect-modifier ranking, allowing direct comparison of
its importance against baseline covariates; it is not, however, a candidate
in the Q2 regional-covariate ranking, where it serves as the response
variable.

\textbf{Top-ranked covariates and overlap recovery.}\\
Figure~\ref{fig:sim-hit} reports identification performance based on
top-1 rankings and overlap recovery for the target covariate under the
observed and unobserved analysis-input cases. The three panels follow the
workflow sequence: the Q2 regional-covariate ranking (Panel~A), the Q3
effect-modifier ranking (Panel~B), and their overlap (Panel~C). 

Panel~A displays the probability that the top-1 covariate in the Q2
regional-covariate ranking is the target, plotted against OR. In the observed
case, once regional imbalance reaches a moderate level
(OR$\,{\geq}\,2$), the true effect modifier $X_{\text{pred}}$ is identified
as the most regionally imbalanced covariate with near-certainty; results are
pooled over $\beta_1/\beta_1^{*}$ because the regional-covariate ranking
depends on the $\mathit{Region}$--covariate association rather than on
treatment effect modification. In the unobserved case under Scenario~2,
proxy covariates, particularly $X_{12}$ which is strongly correlated with
the excluded $X_{14}$, recover this signal with comparable performance at
higher OR, whereas Scenario~1 (no usable proxy) is not evaluable against an
empty truth set.

Panel~B displays the corresponding probability for the Q3 effect-modifier
ranking, plotted against $\beta_1/\beta_1^{*}$. In the observed case, the
probability of correctly identifying $X_{\text{pred}}$ as the top-1 effect
modifier increases steeply with the strength of treatment--covariate
interaction. Results are pooled over OR (black lines), as the effect-modifier
ranking is largely insensitive to regional imbalance which is consistent with the
orthogonal sensitivity of $p_{\mathrm{TEH}}$ observed in Objective~1. The
same pooled presentation applies to unobserved Scenario~2, where proxy
covariates, joined by $\mathit{Region}$ as part of the evaluable target
set, are recovered with attenuated but appreciable probability. Unobserved Scenario~1 is the exception:
here, no strong proxy exists and $\mathit{Region}$ itself absorbs part of the
effect-modification signal, acting as a surrogate for the omitted
$X_{\text{pred}}$ in the pathway
(cf.\ Figure~\ref{fig:Q2-vi}). Because the strength of the
$\mathit{Region}$--$X_{\text{pred}}$ association is governed by OR, the top-1
selection probability of $\mathit{Region}$ now depends on both
$\beta_1/\beta_1^{*}$ and OR; colored lines stratified by OR are therefore
shown for this panel, reaching approximately $0.2$ at
$\beta_1/\beta_1^{*}=2$, OR$\,{=}\,10$.

Panel~C combines both component rankings into the overlap, displaying the
probability that the target covariate appears in
$\mathcal{T}_{\mathrm{Reg}}(5)\cap\mathcal{T}_{\mathrm{EM}}(5)$, plotted
against $\beta_1/\beta_1^{*}$. Solid and dashed lines distinguish
OR$\,{\geq}\,2$ and OR$\,{<}\,2$, respectively. In the observed case,
overlap recovery increases steeply with $\beta_1/\beta_1^{*}$ and reaches
near-certainty for OR$\,{\geq}\,2$ at
$\beta_1/\beta_1^{*}\geq 1.5$. The modest separation between solid and
dashed lines reflects the fact that the Q2 ranking already achieves
near-perfect top-1 recovery at OR$\,{\geq}\,2$ (Panel~A), so that residual
variation in overlap recovery at these imbalance levels is driven primarily
by the power of the effect-modifier ranking (Panel~B). At OR$\,{=}\,1.5$
(dashed), the slightly lower overlap probability traces to a modest reduction
in Q2 performance rather than to a qualitative change in behavior.
In the unobserved case, Scenario~2 retains appreciable overlap recovery
through proxy covariates, while Scenario~1 shows diffuse and low overlap
probabilities, confirming that the overlap set remains uninformative when no
proxy structure is available.

Mapping these patterns to the roadmap (Table~\ref{tab:roadmap}): when both
rankings converge on the true effect modifier in the observed case, the
non-empty overlap directly supports T2 (explained regional heterogeneity). In
the unobserved case, Scenario~2 supports T3 (partial explanation through
measured surrogates), while the diffuse overlap in Scenario~1 leads toward T4
(unexplained heterogeneity), alerting the analyst that observed covariates do
not account for the regional differences detected by $p_{\mathrm{RV}}$.

\textbf{Per-variable selection profiles.}\\
To complement the aggregate top-1 summaries in Figure~\ref{fig:sim-hit}, Figure~\ref{fig:sim-profiles} adopts the same top-5 window used to define the overlap and disaggregates the results to the individual covariate level.
Panel~A displays the probability of appearing in the Q2
regional-covariate set $\mathcal{T}_{\mathrm{Reg}}(5)$, stratified by OR
(rows), as the primary signal for Q2. Panel~B shows the probability
of appearing in the Q3 effect-modifier set $\mathcal{T}_{\mathrm{EM}}(5)$,
stratified by $\beta_1/\beta_1^{*}$ (rows), as the primary signal for
Q3. Panel~C shows the probability of appearing in the overlap
$\mathcal{T}_{\mathrm{Reg}}(5)\cap\mathcal{T}_{\mathrm{EM}}(5)$, stratified
by both OR (rows) and $\beta_1/\beta_1^{*}$ (row facets), as the overlap
requires both signals to be present. All panels are shown separately for the
two scenarios and analysis-input cases.

In the observed case, the true effect modifier $X_{\text{pred}}$ is
consistently selected with the highest probability in both component rankings
and hence dominates the overlap. In Scenario~2, the proxy covariates
$X_{12}$ and $X_9$ also attain non-negligible selection probabilities in the
Q2 ranking (Panel~A; e.g., $X_{12}$ and $X_9$ both reaching approximately
$0.5$--$0.8$ at high OR), reflecting their correlation with the true modifier
$X_{14}$. This co-selection is not a methodological artifact but a natural
consequence of the shared information between correlated covariates, and it
does not prevent the method from correctly prioritizing $X_{14}$ at the top
of both rankings. A notable finding from Panel~B is that $X_{\text{pred}}$
is consistently selected with higher probability than $\mathit{Region}$ in
the effect-modifier ranking, and this separation becomes more pronounced as
$\beta_1/\beta_1^{*}$ increases. For example, in Scenario~2 at the highest
signal level, the ordering is
$X_{14} > X_{12} > \mathit{Region} \approx X_9$. This indicates that the
method attributes treatment effect heterogeneity to the underlying baseline
covariate rather than the regional label, even though both are available as
candidate modifiers, a property that is central to the workflow's goal of
\emph{explaining}, rather than merely detecting, regional differences.

In the unobserved case, the two scenarios diverge markedly. Scenario~1
exhibits low and diffuse selection across all covariates in both rankings,
consistent with the absence of strong proxy structure: no single covariate
inherits a strong association with either $\mathit{Region}$ or the treatment
effect. In Panel~B, $\mathit{Region}$ emerges as the most frequently selected
effect-modifier candidate (reaching approximately $0.3$ at high signal),
reflecting its role as a surrogate for the omitted $X_{11}$. Scenario~2, by
contrast, shows concentration on the proxy covariates, with the two component
rankings exhibiting distinct patterns. In the Q2 regional-covariate ranking
(Panel~A), both $X_{12}$ and $X_9$ attain high selection probabilities
(reaching approximately $0.9$ at high OR), effectively replacing the excluded
$X_{14}$ as the most regionally distinguishing covariates. In the Q3
effect-modifier ranking (Panel~B), $X_{12}$ dominates (selection probability
exceeding $0.8$ at high signal), while $\mathit{Region}$ and $X_9$ are
selected at lower but non-negligible rates (${\approx}\,0.3$), somewhat
higher than their corresponding rates in the observed case
(${\approx}\,0.2$), consistent with these variables partially absorbing the
signal of the excluded modifier. The convergence of $X_{12}$ at the top of
both component rankings explains the appreciable overlap recovery observed in
Figure~\ref{fig:sim-hit}C.

To make the diffuse selection patterns in the no-proxy setting more
interpretable, Supplementary Figure~S2 provides
detailed per-covariate selection probabilities for a representative
high-signal setting (OR$\,{=}\,10$, $\beta_1/\beta_1^{*}=2$), confirming
the proxy dominance under Scenario~2 and revealing that under Scenario~1,
a few weakly correlated covariates attain modestly elevated selection
probabilities---explaining why the diffuse pattern is not perfectly uniform
across covariates.


\section{Discussion}
\label{sec:discussion}


We have proposed a structured, question-driven workflow for exploring
regional treatment effect heterogeneity in MRCTs. Inspired by the
WATCH framework for general TEH
assessment~\cite{Sechidis2025WATCH}, the present workflow addresses
a distinct problem: decomposing observed regional differences into
covariate-level signals that can be interpreted alongside clinical
and regulatory knowledge. The simulation study demonstrated that
the three global evidence summaries ($p_{\mathrm{RV}}$,
$p_{\mathrm{RI}}$, $p_{\mathrm{TEH}}$) are well calibrated under
null-like settings and exhibit sensitivity to their respective
signal components. The overlap-based identification reliably
recovers region-associated effect modifiers when they are observed,
and the methods consistently attribute heterogeneity to the
underlying baseline covariate rather than the regional label itself,
even when both are available as candidate modifiers.


A key motivation for the roadmap (Table~\ref{tab:roadmap}) is that,
in practice, the analyst does not know whether the true effect
modifier is observed, unobserved but recoverable through proxy
covariates, or entirely unobserved with no usable proxy. The
simulation results suggest that the evidence pattern itself can guide this judgment. When $p_{\mathrm{RV}}$ is unremarkable, the
methods support regional consistency~(T1), though this should be
interpreted in light of the regional sample sizes available. When $p_{\mathrm{RV}}$ signals heterogeneity and the overlap contains a clearly dominant covariate that is also clinically interpretable,
the pattern supports an identified candidate factor~(T2). When
$p_{\mathrm{RV}}$ is notable yet both $p_{\mathrm{RI}}$ and
$p_{\mathrm{TEH}}$ remain flat and the overlap is empty or
diffuse, the observed covariates do not account for the regional
differences, pointing toward unexplained heterogeneity~(T4).
Between these extremes, partial explanation~(T3) is supported when the overlap contains covariates that are plausible proxies rather than direct effect modifiers, or when $\mathit{Region}$ itself
ranks highly in the Q3 effect-modifier ranking, suggesting that
regional grouping captures heterogeneity not fully explained by
the observed covariates. These interpretive patterns are intended
as heuristic guides rather than formal decision rules; their value
lies in providing a disciplined starting point for the
multidisciplinary assessment that should accompany any exploratory
analysis of regional heterogeneity.

The simulation distinguished ``true'' modifiers from ``proxies''
because the data-generating mechanism was known. In practice,
however, this distinction is rarely available. A covariate
identified through the overlap may itself be a proxy for an
unmeasured region-associated effect modifier, yet still carry
clinically meaningful information.
As an illustration, consider irinotecan
treatment in multinational oncology trials: UGT1A1 polymorphisms
(e.g., *28, more prevalent in Europeans and Africans; *6, more
prevalent in East Asians) reduce detoxification of the active
metabolite SN-38, increasing toxicity risk, yet are not routinely
genotyped in global trials. Baseline total bilirubin, which is
always recorded, correlates with UGT1A1 enzyme activity and
differs in distribution across populations with different allele
frequencies~\citep{Parodi2008UGT1A1Bilirubin}. If such a trial were analyzed using the proposed
workflow, bilirubin could plausibly enter the Q2/Q3 overlap as a
phenotypic proxy for the unmeasured genetic factor, providing an
actionable signal even without direct genotyping data. The
corresponding conclusion~(T3) appropriately communicates that
heterogeneity is partially but meaningfully explained, while
acknowledging that residual variability may warrant further
investigation.


The contribution of work is the structured question-driven workflow, which is model-agnostic. Various approaches for
evaluating treatment effect heterogeneity exist in the
literatures~\citep{lipkovich2024modern}. The implementation shown in running example and simulation, including DR-learner
for pseudo-outcome construction, conditional random forests for
variable ranking, and permutation-based variable importance profiles, is one instantiation. Alternative approaches at each step could be considered when preserving the workflow logic.

For instance, one may directly assess
whether adjusting for a candidate covariate attenuates the
region-by-treatment interaction, as in the PLATO trial
analysis~\citep{CarrollFleming2013PLATO, dane2019subgroup}. Such an approach provides a decomposition of the region-by-treatment interaction by sequentially assessing the attenuation attributable to each candidate covariate. It addresses Q2 and Q3 jointly,
identifying covariates whose adjustment most reduces the
interaction. Our proposed workflow leverages multivariate, correlation-aware rankings and provides structured evidence
summaries at each node, while the decomposition approach may be more
transparent in settings with few candidate covariates. Regarding
the construction of individual-level pseudo-observations, we used
the DR-learner throughout but verified in preliminary checks that
score-residual-based alternatives~\cite{chen2026comparing} yielded
qualitatively similar results for the continuous endpoint considered
here. For outcomes on model-based scales (e.g., hazard ratios or
odds ratios), score residuals of the treatment effect parameter
from a fitted regression model can serve the same role, enabling
application of the workflow to a broader range of endpoint types. At the variable ranking level, alternatives to
conditional random forests, such as SHAP-based importance or
parametric interaction screening, could be employed. Regarding variable ranking, any method for deriving importance scores from
pseudo-outcomes could be considered, such as SHapley Additive exPlanations (SHAP)
values~\citep{Lundberg2017} as well as those discussed by Hooker et
al.~\citep{Hooker2021}.


Several limitations should be noted. First, the workflow provides a
descriptive decomposition of regional heterogeneity rather than a
causal explanation: the relationship between region and baseline
covariates is associational, and overlap membership does not
establish that a covariate causally mediates the regional
difference. Clinical and scientific judgment remains essential for
interpreting the results. Second, the simulation considered a
binary region indicator and a continuous endpoint; MRCTs typically
involve multiple regions with varying sample sizes, and endpoints
may be binary or time-to-event. The simulated regional prevalence of 20\% is representative of
moderately sized subpopulations; smaller regions (e.g., 5--10\%)
would further reduce power for the identification of
region-associated effect modifiers. Methods for regional sample size determination and consistency
assessment under such designs have been developed
separately~\citep{RenXu2026consistency, QingRenXu2025twoMRCTs,
AdallXu2021BayesShrinkage}; integrating these design-stage
considerations with the post-hoc exploration proposed here
represents a direction for future work. While the workflow can in principle
be applied to these settings, for example, using risk differences
for binary endpoints or score residuals for hazard
ratios~\cite{chen2026comparing}, the behavior of the variable
importance rankings and overlap identification under these
conditions has not been systematically evaluated. 

Several directions merit further investigation. Extending the
workflow to multi-region settings with hierarchical structure (e.g.,
countries nested within regions) would increase practical relevance.
Mediation-inspired decomposition or covariate standardization
approaches could complement the current descriptive framework by
quantifying the proportion of regional heterogeneity attributable
to specific covariates, though their properties in the MRCT setting
require careful study. The proposed workflow aligns with regulatory
expectations across multiple frameworks: ICH~E17~\citep{ICH2017}, the FDA implementation guidance~~\citep{FDA2018E17Guidance},
the Japanese MHLW guidance~\citep{ministry2007basic}, and the CDE
guideline on benefit--risk assessment based on MRCT
data~\citep{CDE2026MRCT} all emphasize structured investigation of
regional inconsistencies without prescribing fixed statistical
thresholds; integration of Q1--Q4 outputs into these regulatory
workflows represents a natural extension. More broadly, the
structured, question-driven approach advocated here aims to reduce
analytic variability in post-hoc explorations of regional
heterogeneity, supporting proportionate and transparent
interpretation of MRCT data.

\section{Disclosure statement}\label{disclosure-statement}

Cong Zhang, Kostas Sechidis, Xiaoni Liu, Sophie Sun, Yao Chen, Shuhei Kaneko and Björn Bornkamp are employees of Novartis. The remaining authors declare no conflicts of interest.



\phantomsection\label{supplementary-material}
\bigskip

\begin{center}
{\large\bf SUPPLEMENTARY MATERIAL}
\end{center}

\begin{description}

\item[Supplementary Figures:]
Additional simulation results including calibration of global
evidence summaries under the unobserved analysis case
(Supplementary Figure~S1), detailed per-covariate selection probabilities at
a representative high-signal setting (Figure~S2), and proxy
correlation structure (Table~S1). (PDF file)

\end{description}

\newpage
\begin{center}
{\large\bf SUPPLEMENTARY MATERIAL}

\bigskip

A Workflow for Evaluating Regional Treatment Effect
Heterogeneity in Multi-Regional Clinical Trials
\end{center}

\bigskip

\setcounter{figure}{0}
\setcounter{table}{0}
\renewcommand{\thefigure}{S\arabic{figure}}
\renewcommand{\thetable}{S\arabic{table}}

\subsection*{S1. Calibration of global evidence summaries under
the unobserved case}

\begin{figure}[htbp]
  \centering
  \includegraphics[width=\textwidth]{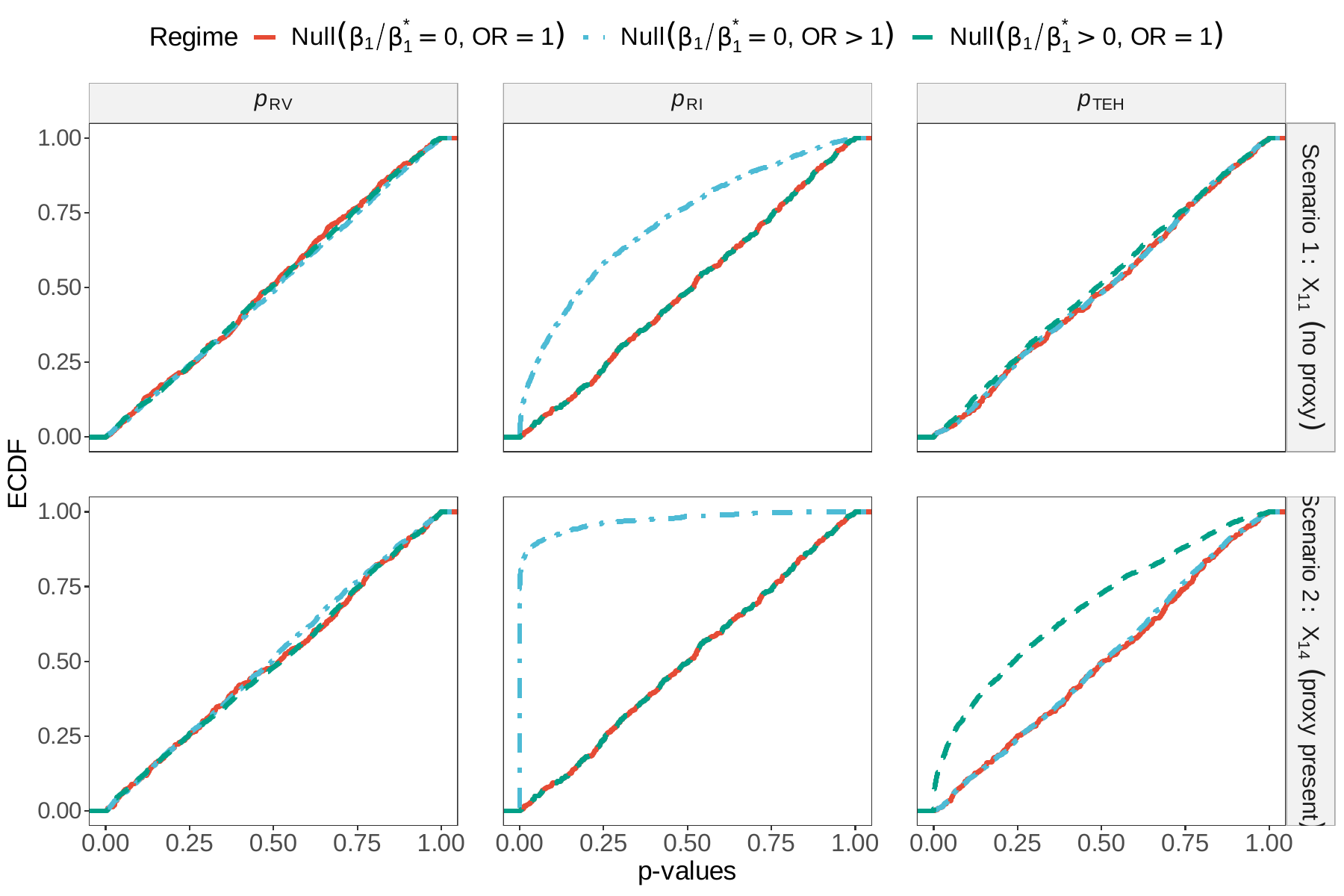}
  \caption{\textbf{Calibration of global evidence summaries under
  the unobserved case.} Empirical CDFs of $p_{\mathrm{RV}}$,
  $p_{\mathrm{RI}}$, and $p_{\mathrm{TEH}}$ under three null-like
  settings, shown for the unobserved analysis case. Layout as in
  Figure~4 (main text). The diagonal line indicates the
  Uniform$(0,1)$ reference. Rows correspond to Scenarios~1--2.}
  \label{fig:supp-ecdf}
\end{figure}

Figure~\ref{fig:supp-ecdf} replicates the calibration assessment
of Figure~4 (main text) under the unobserved analysis case.
Results are consistent with the observed case: under complete
homogeneity ($\beta_1/\beta_1^{*}=0$, OR$=1$), all three
summaries are close to Uniform$(0,1)$. Under regional imbalance
without TEH ($\beta_1/\beta_1^{*}=0$, OR$>1$), only
$p_{\mathrm{RI}}$ deviates toward small values. Under TEH without
regional imbalance ($\beta_1/\beta_1^{*}>0$, OR$=1$), only
$p_{\mathrm{TEH}}$ deviates. In Scenario~1 (no usable proxy),
$p_{\mathrm{RI}}$ and $p_{\mathrm{TEH}}$ show reduced sensitivity
compared to the observed case, reflecting the absence of the true
region-associated covariate from the analysis set. In Scenario~2,
where the strong proxy $X_{12}$ is available, $p_{\mathrm{RI}}$
retains sensitivity comparable to the observed case.

\subsection*{S2. Detailed per-covariate selection probabilities}

\begin{figure}[htbp]
  \centering
  \includegraphics[width=\textwidth]{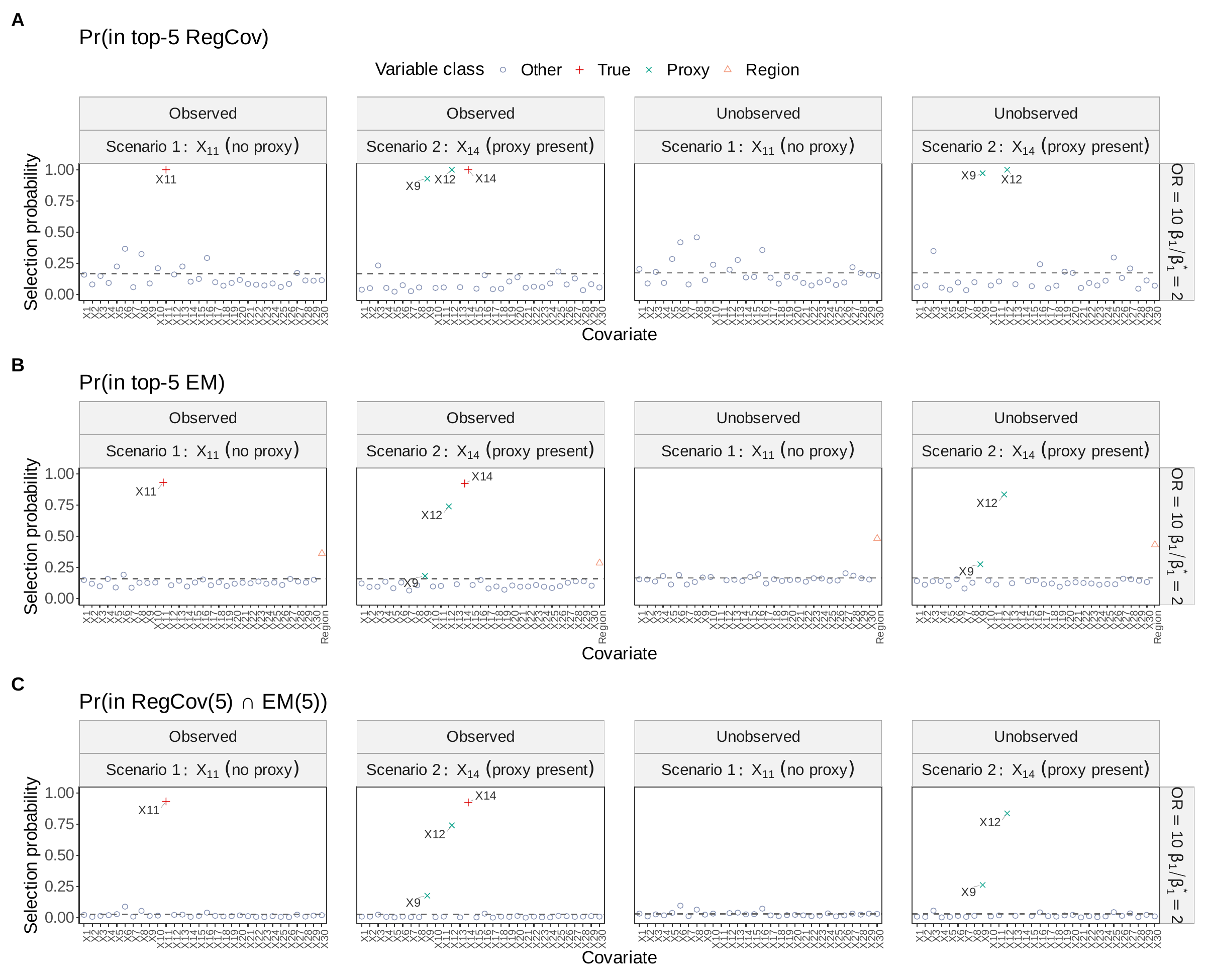}
  \caption{\textbf{Detailed per-covariate selection probabilities
  at OR$=10$, $\beta_1/\beta_1^{*}=2$.} Per-covariate top-5
  selection probabilities for (A)~the overlap set
  $\mathcal{T}_{\mathrm{Reg}}(5)\cap\mathcal{T}_{\mathrm{EM}}(5)$,
  (B)~the Q2 regional-covariate set
  $\mathcal{T}_{\mathrm{Reg}}(5)$, and (C)~the Q3 effect-modifier
  set $\mathcal{T}_{\mathrm{EM}}(5)$, across observed and
  unobserved cases for Scenarios~1--2. Marker color indicates
  variable class: true effect modifier (red), proxy covariates
  (green), $\mathit{Region}$ (salmon), and other covariates (grey). }
  \label{fig:supp-profiles}
\end{figure}

Figure~\ref{fig:supp-profiles} provides detailed per-covariate
selection probabilities for a representative high-signal setting
(OR$\,{=}\,10$, $\beta_1/\beta_1^{*}=2$). This complements
Figure~7 (main text) by displaying all individual covariates,
confirming proxy dominance in Scenario~2 (unobserved case) and
revealing that under Scenario~1 (unobserved case), a few weakly
correlated covariates attain modestly elevated selection
probabilities without any single variable reaching clear
prominence.

\subsection*{S3. Proxy correlation structure}

Table~\ref{tab:supp-corr} reports pairwise latent Gaussian
correlations between the predictive covariates ($X_{11}$,
$X_{14}$) and the remaining baseline covariates, estimated via
the \texttt{latentcor} R package. These correlations support the
interpretation of proxy-driven patterns in the simulation results
(Section~4 of the main text).

\begin{table}[htbp]
  \centering
  \caption{\textbf{Pairwise latent correlations between predictive
  covariates and remaining baseline variables.} The five strongest
  absolute correlations are shown for each scenario (mean across
  500 replications).}
  \label{tab:supp-corr}
  \begin{tabular}{llcc}
    \hline
    Scenario & Covariate pair & $|\hat{\rho}|$ & Role \\
    \hline
    1 & $(X_{11}, X_{j})$ for all $j$ & $\leq 0.17$ &
      No usable proxy \\
    \hline
    2 & $(X_{14}, X_{12})$ & 0.86 & Strong proxy \\
      & $(X_{14}, X_{9})$  & 0.41 & Moderate proxy \\
      & $(X_{14}, X_{j})$ for others & $\leq 0.20$ & Weak \\
    \hline
  \end{tabular}
\end{table}



  \bibliography{bibliography.bib}

\end{document}